\documentclass[a4paper,final]{iopart}
\usepackage{iopams}
\usepackage{setstack}
\usepackage{times}
\usepackage{mathbbol}
\def\onehalf{{\textstyle\frac{1}{2}}}
\def\quarter{{\textstyle\frac{1}{4}}}
\def\HC{e}
\def\QB{Q}

\def\RB{\mathbb{R}}
\def\d{\mathrm{d}}
\def\vecq{\bi{q}}
\def\vecp{\bi{p}}
\def\vecA{\bi{A}}

\def\vecP{\bi{P}}
\providecommand{\Eins}{\mathbf{1}}
\begin{document}
\title[Hamiltonian dynamics on $T^{*}\QB_{1}$]
{Hamiltonian dynamics on the symplectic extended phase
space for autonomous and non-autonomous systems}
\author{J\"urgen Struckmeier}
\address{Frankfurt Institute for Advanced Studies (FIAS),\newline
Ruth-Moufang-Str.~1, 60438~Frankfurt am Main, Germany}
\ead{struckmeier@fias.uni-frankfurt.de}
\begin{abstract}
We will present a constrained description of Hamiltonian dynamics on the ``symplectic extended phase space'' that is analogous
to that of a time-\underline{in}dependent Hamiltonian system on the conventional symplectic phase space.
The energy constraint and the pertaining extended symplectic structure will be derived from a generalized formulation of Hamilton's variational principle.
An extended Hamiltonian $H_{1}=k(H-\HC)$ generates the physical dynamics on the constraint surface, with $k$ specifying the parameterization of time.
The extended canonical transformation theory naturally permits transformations that also map the time scales of original
and destination system while preserving the form of the canonical equations.
For generating functions without explicit dependence on the common evolution parameter, the extended Hamiltonian transforms as a scalar function.
The Lorentz transformation and prescribed time scaling transformations will be shown to represent particular canonical transformations in the symplectic extended phase space.
Coordinate-dependent time rescalings and the constrained Kustaanheimo-Stiefel mapping will be distinguished from invertible canonical coordinate transformations.
Furthermore, an ``extended'' generating function that maps explicitly time-dependent Hamiltonians into time-independent ones
will be presented for the time-dependent damped harmonic oscillator and for an explicitly characterized class of time-dependent potentials.
For a general potential, a related auxiliary system yields a trajectory-dependent reconstruction identity rather than a universal linear evolution law.
In the appendix, the singular Legendre transformation of the extended Lagrangian $L_{1}$ will be shown to yield the energy constraint without uniquely fixing its Hamiltonian extension away from the constraint surface.
\end{abstract}
\def\published#1{\vspace{28pt plus 10pt minus 18pt}
     \noindent{\small\rm #1\par}}
\published{Revised manuscript based on J.~Phys.~A: Math.~Gen.~{\bf 38} (2005) 1257--1278}
\section{\label{sec:intro}Introduction}
The modern description of time-\underline{in}dependent
Hamiltonian systems on symplectic manifolds is well
established (e.g.\ Abraham and Marsden 1978, Arnold 1989,
Jos\'e and Saletan 1998, Marsden and Ratiu 1999, Frankel 2001).
If the Hamiltonian $H$ is explicitly time dependent, the instantaneous phase space $T^{*}\QB$ remains symplectic, but its Hamiltonian vector field depends on time.
Alternatively, one may work on the odd-dimensional evolution space $T^{*}\QB\times\RB$ with the closed presymplectic $2$-form $\omega_{H}=\omega-\d H\wedge\d t$ (Abraham and Marsden 1978, Marsden and Ratiu 1999, Frankel 2001).
The conventional time-dependent canonical transformation theory keeps $t$ as the common independent variable of original and destination system.
Transformations that change the time coordinate, such as Lorentz transformations and the time rescalings used in celestial mechanics, are more naturally formulated after adjoining its conjugate momentum.
More general transformations can also be studied in presymplectic geometry, so this distinction concerns the chosen formalism rather than a general impossibility within that geometry.
Regularization transformations dating back to L.~Euler (Siegel and Moser 1971) provide an important application.
Suitable transformations of non-linear, explicitly time-dependent Hamiltonian systems into time-independent systems can likewise involve non-trivial mappings of the time scales (Struckmeier and Riedel 2001).
Such mappings are not necessary for every removal of explicit time dependence.

Various approaches were made to describe these transformations
within the context of a generalized canonical transformation
theory (Lanczos 1949, Synge 1960, Szebehely 1967, Kuwabara 1984, Asorey \etal
1983, Cari\~{n}ena \etal 1987, Cari\~{n}ena \etal 1988).
The underlying idea is to develop a generalized Hamiltonian
formalism on a ``symplectic extended phase space'' in analogy
to the symplectic description on the conventional phase space
of an autonomous Hamiltonian system ($\partial H/\partial t=0$).
Specifically, in the extended formalism,  the time $t$ is treated
as an ordinary canonical function $t(s)\equiv q^{n+1}(s)$ of a
new superordinated system evolution parameter, $s$.
Its canonically conjugate function $p_{n+1}(s)$ will constitute
the additional coordinate that renders the carrier manifold
even-dimensional --- and hence eligible for a symplectic description.
The dynamics of the given system is then represented on an energy-constraint surface by an extended Hamiltonian $H_{1}$.
A prescribed phase-space scaling factor without explicit $s$ dependence gives $\partial H_{1}/\partial s=0$.
As a consequence, all properties of Hamiltonian systems on
symplectic manifolds can then be similarly reformulated
within the extended Hamiltonian formalism.
For example, time-dependent symmetries of explicitly
time-dependent Hamiltonian systems can be treated
like usual symmetries of autonomous Hamiltonian systems.
Moreover, it is possible to define canonical transformations within the symplectic extended phase space that are more general than conventional time-preserving transformations.
This formalism will then naturally permit generating
functions of extended canonical transformations that also
define a non-trivial mapping of the time-scales of the
original and the destination systems.
As in the conventional canonical transformation theory, preservation of the canonical equations does not require numerical preservation of the Hamiltonian.
An $s$-independent generating function yields the passive transformation law $H_{1}^{\prime}(z^{\prime})=H_{1}(z)$, where $z$ denotes an extended phase-space point.
An explicitly $s$-dependent generating function contributes its partial $s$ derivative to the transformed Hamiltonian.

Following the pioneering works of C.~Lanczos (Lanczos 1949, p~189) and J.~L.~Synge (Synge 1960, p~143), the energy constraint is commonly written as $H_{\mathrm{LS}}=H-\HC=0$ (e.g.\ Stiefel and Scheifele 1971, Thirring 1977, Asorey \etal 1983, Kuwabara 1984, Lichtenberg and Lieberman 1992, Stump 1998, Tsiganov 2000, Struckmeier and Riedel 2002a).
Here $\HC$ is an independent coordinate in the extended phase space and equals the value of $H$ on the physical constraint surface.
The choice $H_{\mathrm{LS}}=H-\HC$ is a valid extended Hamiltonian and fixes $\d t/\d s=1$.
After a canonical transformation that changes time, the transformed Hamiltonian need not retain this particular normalization of the constraint.

Other approaches (e.g.\ Gotay 1982, Cari\~{n}ena \etal 1987, Cari\~{n}ena \etal 1988) use a coisotropic embedding of the presymplectic evolution space into the symplectic extended phase space.
Hamiltonians of the form $H_{\mathrm C}=f(H-\HC)$, with $f$ a smooth nonvanishing phase-space function, are compatible with a Hamiltonian variational principle.
On the constraint surface they generate the same physical trajectories with a different parameterization.
Derivatives of $f$ contribute away from that surface and must not be omitted from unrestricted canonical equations.

With this article, we aim to present these features in a unified symplectic description that makes the role of the energy constraint and the time parameterization explicit.
The derivation from Hamilton's principle will be the starting point of our analysis in Section~\ref{sec:principle}.
The extended Hamiltonian $H_{1}=k(H-\HC)$, with $\d t/\d s=k$ on the physical surface, coincides with the Hamiltonian of the Poincar\'e time transformation (Siegel and Moser 1971, p~35).
For a chosen smooth phase-space function $k$, Section~\ref{sec:caneq} will establish the corresponding canonical equations both on and away from the constraint surface.
Different extensions away from that surface can represent the same constrained dynamics.

In Section~\ref{sec:extended}, we will derive the formalism of extended canonical trans\-for\-mations within the symplectic extended phase space.
These transformations preserve the form of the canonical equations, with the Hamiltonian transformation law determined by the generating function.
In this description, the evolution parameter $s$ serves
as the independent variable that is common to both the
original and the destination system.
In this respect, the new evolution parameter $s$ plays exactly
the role of the time $t$ in the conventional canonical
transformation theory.
The generalized formulation of Hamilton's variational principle
thus establishes the basis for the definition of {\em extended
generating functions\/} for {\em finite\/} canonical transformations
that make it possible to relate both systems at {\em different\/}
instants of their respective time scales, $t(s)$ and $t^{\prime}(s)$.
We will furthermore formulate the extended version of Liouville's
theorem that applies to the symplectic extended phase space.
In addition, the restrictions will be worked out that are to be
imposed on the functional structure of extended generating
functions in order for the transformed time $t^{\prime}(s)$
to sustain the meaning of $t(s)$ as a common parameter for
all canonical variables $p_{i}^{\prime}$ and $q^{\prime\,i}$.

As a first example of an extended canonical transformation,
we will show in Section~\ref{sec:lorentz} that the Lorentz
transformation can be formulated as a particular canonical
transformation in the extended phase space.
Its generating function does not explicitly depend on $s$.
The resulting passive scalar transformation law for $H_{1}$ will be distinguished from the Lorentz invariance of a particular physical constraint.

In celestial mechanics, regularization transformations
of many-body systems are known to require a replacement
of the physical time $t$ by a ``fictitious'' time $t^{\prime}$.
We will show in Section~\ref{sec:tscal} how a prescribed time-only scaling is implemented as a finite extended canonical transformation.
Coordinate-dependent rescalings used in regularization will instead be formulated through a scaled energy constraint.
The Kustaanheimo-Stiefel example in Section~\ref{sec:ks} additionally requires a bilinear constraint and reduction by its gauge symmetry.

In Section~\ref{sec:hosci}, an extended generating function
will be presented that defines a canonical mapping of a
general class of explicitly time-dependent Hamiltonian
systems into time-\underline{in}dependent ones.
Demanding the transformed system to be autonomous
then determines the time correlation of both systems.
For the simple but important case of a time-dependent
damped harmonic oscillator, the extended canonical transformation
yields the well known invariant given by Leach (Leach 1978).
For the non-linear and explicitly time-dependent Hamiltonian systems treated in Section~\ref{sec:genosci}, we will characterize the potentials for which the displayed generator yields an autonomous destination system.
A related auxiliary linear system can be constructed along a general trajectory.
It reconstructs the initial values of the energy $\HC(t)$ and the scalar products $\vecq(t)\vecp(t)$ and $\vecq^{2}(t)$ through a trajectory-dependent transfer matrix.

In Appendix~\ref{sec:appa}, the parameterization of time will be reviewed for Lagrange's formulation of dynamics.
The singular Legendre transformation of the extended Lagrangian $L_{1}$ yields the same energy constraint as the variational construction in Section~\ref{sec:principle}.
It does not uniquely fix an extended Hamiltonian away from that constraint surface.
\section{\label{sec:general}Hamiltonian formalism
in the symplectic extended phase space}
\subsection{\label{sec:principle}Hamilton's variational principle,
extended Hamiltonian}
We consider an explicitly time-dependent Hamiltonian $H$ on the evolution space $T^{*}\QB\times\RB$ with its closed, degenerate $2$-form $\omega_{H}=\omega-\d H\wedge\d t$.
Herein, $\omega$ stands for a symplectic, i.e.\ closed and non-degenerate $2$-form on $T^{*}\QB$.
The canonical $1$-form $\lambda$ satisfies $\d\lambda=\omega$ in the sign convention used here.
Hamilton's variational principle states that the actual system trajectory $C_{0}\subset T^{*}\QB\times\RB$ is a critical point of the action $\mathfrak{S}$, defined as the integral of the Poincar\'e-Cartan $1$-form $\lambda-H\d t$,
\begin{equation}\label{la-principle}
\mathfrak{S}(C)=\int_{C}(\lambda-H\d t)\,,\qquad\d\mathfrak{S}(C_{0})=0\,.
\end{equation}
This $1$-form is not assumed to be a contact form.
In canonical coordinates, we have $\lambda=\vecp\d\vecq$, with $(\vecq,\vecp)\in T^{*}\QB$ the pair of $n$-component vectors and covectors.
The endpoint configurations and endpoint times are held fixed in the variational problem.

As the basis for a symplectic description, we treat time $t=t(s)\equiv q^{n+1}(s)$ as an ordinary canonical variable depending on a newly introduced system evolution parameter $s$.
The extended configuration manifold is the product $\QB_{1}=\QB\times\RB$, with local coordinates $\vecq_{1}=(\vecq,t)$.
Its cotangent bundle $T^{*}\QB_{1}$ is naturally symplectic.
The additional canonical momentum $p_{n+1}$ is independent of the spatial momenta and of $t$ before a constraint is imposed.
With $\vecp_{1}=(\vecp,p_{n+1})$, the canonical $1$-form is $\Lambda=\vecp_{1}\d\vecq_{1}$.

Introduce an independent multiplier $k(s)$ and define the extended variational principle by
\begin{equation}\label{principle}
\mathfrak{S}_{1}=\int_{s_{1}}^{s_{2}}
\left[\sum_{i=1}^{n+1}p_i\frac{\d q^i}{\d s}-H_{1}\right]\d s\,,
\qquad\delta\mathfrak{S}_{1}=0\,,
\end{equation}
where
\begin{equation}\label{h1-def}
H_{1}=k(s)\left(H+p_{n+1}\right)\,.
\end{equation}
Variation includes the multiplier as well as the canonical variables.
Variation of $k$ gives $H+p_{n+1}=0$, while variation of $p_{n+1}$ gives $\d t/\d s=k$.
Thus, the latter relation is an equation obtained by variation and is not substituted into the action before variation.
On intervals where $k\ne0$, the remaining equations are equivalent to the conventional canonical equations.

We use $\HC=-p_{n+1}$ as the independent energy coordinate and denote the physical constraint surface by $\Sigma$,
\begin{equation}\label{hamid}
\mathcal C(\vecq,\vecp,t,\HC)=H(\vecq,\vecp,t)-\HC\,,\qquad
\Sigma=\{\mathcal C=0\}\,.
\end{equation}
Along a physical trajectory, $\HC(s)=H(\vecq(s),\vecp(s),t(s))$.
If $H$ represents the system's energy, $\HC(s)$ gives its instantaneous energy content.
Away from $\Sigma$, $\HC$ remains an independent canonical coordinate.
As will be shown in Appendix~\ref{sec:appb}, the same constraint follows from the singular Legendre transformation of the extended Lagrangian $L_{1}$.

The map from the evolution space into the extended phase space is the embedding
\begin{displaymath}
\psi(\vecq,\vecp,t)=(\vecq,\vecp,t,H(\vecq,\vecp,t))\,.
\end{displaymath}
Its image is $\Sigma$, and its inverse is defined on $\Sigma$, not on the entire extended phase space.
The multiplier formulation gives
\begin{equation}\label{h1-1fdef}
H_{1}\d s=k(s)(H-\HC)\d s\,,\qquad
\frac{\d t}{\d s}=k(s)\quad\hbox{on solutions}\,.
\end{equation}
Eliminating the constrained pair in the action recovers~\eref{la-principle}.
This equivalence fixes the constraint but not a unique Hamiltonian extension away from $\Sigma$.

After choosing a parameterization, one may instead prescribe a smooth phase-space function $k(\vecq,\vecp,t,\HC)$ that is nonzero near the part of $\Sigma$ under consideration.
The resulting autonomous Hamiltonian is
\begin{equation}\label{ham1}
H_{1}(\vecq,\vecp,t,\HC)=k(\vecq,\vecp,t,\HC)\,
\big[H(\vecq,\vecp,t)-\HC\big]\,.
\end{equation}
Its physical initial data are restricted to $\Sigma$.
The relation $\d t/\d s=k$ then holds on $\Sigma$.
The distinction between the independent multiplier $k(s)$ and a prescribed phase-space scaling function $k$ will be maintained below.

With the additional pair $(q^{n+1},p_{n+1})=(t,-\HC)$, the extended symplectic structure is
\begin{equation}\label{omega1}
\Omega=\d\Lambda=\sum_{i=1}^{n+1}\d p_{i}\wedge\d q^{i}=
\sum_{i=1}^{n}\d p_{i}\wedge\d q^{i}-\d\HC\wedge\d t\,.
\end{equation}
Its pull-back by $\psi$ is $\omega_{H}$.
The form of~\eref{ham1} is that of Poincar\'e's time transformation (Siegel and Moser 1971, p~35).
The choice $k=1$ gives $H_{\mathrm{LS}}=H-\HC$ and the parameterization $t=s+\mathrm{const}$.
A phase-space-dependent nonvanishing $k$ is equally admissible and changes the rate of traversal of the same physical trajectories.
This includes Hamiltonians $H_{\mathrm C}=f(H-\HC)$ discussed in the geometric literature (Cari\~{n}ena \etal 1987, Cari\~{n}ena \etal 1988).

The function $H_{1}$ vanishes on $\Sigma$ but not identically on $T^{*}\QB_{1}$.
Its ambient derivatives can therefore generate nontrivial motion on that surface.
This is different from the vanishing Legendre energy of the homogeneous extended Lagrangian on the image of its singular Legendre map.
Even after choosing $k$, adding $(H-\HC)^{2}R$ with $R$ smooth leaves the Hamiltonian vector field unchanged on $\Sigma$.
The physical constrained dynamics therefore does not uniquely determine the off-surface extension.

In the Lagrangian description, $\d t/\d s$ is an independent velocity in the argument list of $L_{1}$.
In the Hamiltonian description, its arbitrariness is represented by the multiplier until a parameterization has been chosen.
Zeros of a scaling factor require separate treatment, since they destroy the local invertibility of the time parameterization.
\subsection{\label{sec:caneq}Canonical equations, Poisson brackets}
For a prescribed phase-space scaling function $k$, the canonical coordinate representation $\Lambda=\vecp_{1}\d\vecq_{1}$ gives the variational integral
\begin{equation}\label{principle1}
\delta\int_{s_{1}}^{s_{2}}\left[ \sum_{i=1}^{n+1}p_{i}(s)
\frac{\d q^{i}(s)}{\d s}-H_{1}\big(\vecq_{1}(s),\vecp_{1}(s)\big)
\right]\d s=0\,.
\end{equation}
The constraint $\mathcal C=0$ is imposed on the initial data in this parameter-fixed action.
The canonical variables are varied independently away from the constraint surface.
The representation formally agrees with the conventional variational principle for a time-independent Hamiltonian.
Therefore, the critical path within the extended phase-space
$(\vecq_{1}(s),\vecp_{1}(s))\subset T^{*}\QB_{1}$ is constituted
by the solution of the extended set of canonical equations
\begin{equation}\label{eqmo1}
\frac{\d q^{i}}{\d s}= \frac{\partial H_{1}}{\partial p_{i}}\,,\quad
\frac{\d p_{i}}{\d s}=-\frac{\partial H_{1}}{\partial q^{i}}\,,\quad
i=1,\ldots,n+1\,.
\end{equation}
In contrast to the total time derivative of the original
Hamiltonian $H$, the total $s$ derivative of $H_{1}$
obviously vanishes identically by virtue of
Equations~\eref{eqmo1}, giving $\d H_{1}/\d s\equiv0$.
Thus, the extended Hamiltonian $H_{1}(\vecq_{1},\vecp_{1})$
from Equation~\eref{ham1} formally converts any given non-autonomous
system $H(\vecq,\vecp,t)$ into an autonomous system in $T^{*}\QB_{1}$.
For a phase-space-dependent $k$, the unrestricted canonical equations are
\begin{eqnarray*}
\frac{\d q^i}{\d s}&=&k\frac{\partial H}{\partial p_i}
+\mathcal C\frac{\partial k}{\partial p_i}\,,\qquad
\frac{\d p_i}{\d s}=-k\frac{\partial H}{\partial q^i}
-\mathcal C\frac{\partial k}{\partial q^i}\,,\\
\frac{\d t}{\d s}&=&k-\mathcal C\frac{\partial k}{\partial\HC}\,,\qquad
\frac{\d\HC}{\d s}=k\frac{\partial H}{\partial t}
+\mathcal C\frac{\partial k}{\partial t}\,.
\end{eqnarray*}
On $\Sigma$, these reduce to
\begin{equation}\label{can1a}
\frac{\d q^{i}}{\d s}=k\,\frac{\partial H}{\partial p_{i}}
\,,\quad\frac{\d p_{i}}{\d s}=-k\,
\frac{\partial H}{\partial q^{i}}\,,\quad
\frac{\d t}{\d s}=k\,,\quad
\frac{\d\HC}{\d s}=k\,\frac{\partial H}{\partial t}\,.
\end{equation}
The leftmost two equations are simply the conventional canonical
equations with $s$ the independent variable instead of $t$.
This shows that physical stationary paths of~\eref{principle} project onto the stationary paths of~\eref{la-principle}, as required.
The rightmost equation from~\eref{can1a} states that the
partial time derivative of $H$ now constitutes a regular
canonical equation --- the equation of motion for $\HC(s)$.
The conjugate equation for $t(s)$ specifies the chosen rate of time evolution.
In the independent-multiplier formulation this rate remains arbitrary until a parameterization is fixed.
For a prescribed phase-space function $k$, it is determined along the chosen trajectory.
If the multiplier depends explicitly on $s$, then $\d H_{1}/\d s=(\partial k/\partial s)\mathcal C$.
Thus autonomy holds throughout the extended phase space for the phase-space choice~\eref{ham1}, but for a general $s$-dependent multiplier the derivative vanishes only on the physical surface.

The equivalence with the original equations requires $k\ne0$ locally.
A negative $k$ reverses the orientation of the parameter along a trajectory.
This is not the same as a physical time-reversal symmetry, which need not be present for an arbitrary Hamiltonian.
A zero of $k$ is a singular limit of the clock transformation rather than an ordinary invertible reparameterization.

The choice $H_{\mathrm{LS}}=H-\HC$ gives $\d t/\d s=1$ and is entirely consistent with treating $t$ as a canonical coordinate.
After a transformation that changes time, one must transform the whole Hamiltonian rather than impose the same unit normalization of both clocks.
The factor $k$ is useful for displaying this freedom explicitly, not for repairing a failure of the canonical equations for $H_{\mathrm{LS}}$.

Obviously, the vector analysis operations on $T^{*}\QB_{1}$ with
its symplectic $2$-form $\Omega$ from Equation~\eref{omega1} are
analogous to the corresponding operations on $(T^{*}\QB,\omega)$.
Let $F:T^{*}\QB_{1}\to\RB$ denote a differentiable function.
We can associate to $F$ a unique vector field $X_{F}$
on $T^{*}\QB_{1}$ by means of the interior product
\begin{displaymath}
X_{F}\lrcorner\,\Omega=-\d F\,.
\end{displaymath}
In canonical coordinate description, this means that we have
along the integral curves of $X_{F}$
\begin{displaymath}
\frac{\d q^{i}}{\d s}=\frac{\partial F}{\partial p_{i}}\,,\quad
\frac{\d p_{i}}{\d s}=-\frac{\partial F}{\partial q^{i}}\,,\quad
i=1,\ldots,n+1\,.
\end{displaymath}
Identifying $F$ with the extended Hamiltonian $H_{1}$,
the restriction of its Hamiltonian vector field to the physical surface is
\begin{equation}\label{dynfield1}
X_{H_{1}}\big|_{\Sigma}=\left.k\,\left[
\sum_{i=1}^{n}\left(\frac{\partial H}{\partial p_{i}}
\frac{\partial}{\partial q^{i}}-\frac{\partial H}{\partial q^{i}}
\frac{\partial}{\partial p_{i}}\right)+\frac{\partial H}{\partial t}
\frac{\partial}{\partial\HC}+\frac{\partial}{\partial t}\right]\right|_{\Sigma}\,,
\end{equation}
with the full off-surface expression $X_{H_{1}}=kX_{\mathcal C}+\mathcal C X_k$.
In particular,
\begin{displaymath}
X_{H_{1}}\lrcorner\,\Omega=-\d H_{1}
=-k(\d H-\d\HC)-\mathcal C\d k\,.
\end{displaymath}
The physical surface $\Sigma$, locally given by $H_{1}=0$ wherever $k\ne0$, is invariant under the flow.
Indeed, $\{\mathcal C,H_{1}\}_{e}=\{\mathcal C,k\}_{e}\mathcal C$, so initial data on $\Sigma$ remain there.
The restriction of $X_{H_{1}}$ to the presymplectic manifold
$T^{*}\QB\times\RB$ then yields the scaled representation
$k\tilde{X}_{H}$ of the vector field $\tilde{X}_{H}$
(Abraham and Marsden 1978, p~376) that describes the dynamics of a
time-dependent Hamiltonian system $(T^{*}\QB\times\RB,\omega_{H},H)$.

Let $F$ and $G$ now denote two differentiable functions on
$T^{*}\QB_{1}$, with $X_{F}$ and $X_{G}$ the related
dynamical vector fields.
The extended symplectic $2$-form induces a corresponding
extended Poisson bracket $\{.,.\}_{e}$ via
\begin{equation}\label{extpb}\fl
-\Omega(X_{F},X_{G})\equiv\{F,G\}_{e}=\sum_{i=1}^{n}\left(
\frac{\partial F}{\partial q^{i}}\frac{\partial G}{\partial p_{i}}-
\frac{\partial F}{\partial p_{i}}\frac{\partial G}{\partial q^{i}}\right)-
\frac{\partial F}{\partial t}\frac{\partial G}{\partial\HC}+
\frac{\partial F}{\partial\HC}\frac{\partial G}{\partial t}\,.
\end{equation}
As is easily verified, the fundamental Poisson brackets are
\begin{displaymath}
\{q^{i},q^{j}\}_{e}=0\,\quad
\{p_{i},p_{j}\}_{e}=0\,\quad
\{q^{i},p_{j}\}_{e}=\delta^{i}_{j}\,\quad i,j=1,\ldots n+1\,.
\end{displaymath}
The extended set of canonical equations~\eref{eqmo1} yields
for differentiable functions on $T^{*}\QB_{1}$, hence
functions $F$ that do not explicitly depend on $s$
\begin{displaymath}
\{F,H_{1}\}_{e}=\frac{\d F}{\d s}\,.
\end{displaymath}
The Poisson bracket representations of the canonical equations are
thus obtained as
\begin{displaymath}
\{p_{i},H_{1}\}_{e}=\frac{\d p_{i}}{\d s}\,,\quad
\{q^{i},H_{1}\}_{e}=\frac{\d q^{i}}{\d s}\,,\quad i=1,\ldots n+1\,.
\end{displaymath}
On $\Sigma$, these equations reproduce the conventional canonical equations expressed in the evolution parameter $s$.

From Equation~\eref{extpb}, we easily confirm that the extended
$2$-form $\Omega$ is non-degenerate.
If $\{F,G\}_{e}=0$ for all differentiable $G$, then $\d F=0$ and $F$ is locally constant.
Equivalently, the contraction map $X\mapsto X\lrcorner\,\Omega$ has trivial kernel.
Consequently, the extended Hamiltonian vector field $X_{H_{1}}$
from Equation~\eref{dynfield1} is uniquely determined by the extended
Hamiltonian $H_{1}$.
This establishes the complete analogy of our extended
description of explicitly time-dependent Hamiltonian systems
$(T^{*}\QB_{1},\Omega,H_{1})$ with the conventional
description of time-\underline{in}dependent
Hamiltonian systems $(T^{*}\QB,\omega,H)$.
\subsection{\label{sec:extended}Canonical transformations,
Liouville's theorem}
By definition, the subset of diffeomorphisms
$\phi:T^{*}\QB_{1}\to T^{*}\QB_{1}$ that preserve the
symplectic structure $\Omega$ are referred to as
symplectic, or, synonymously, as canonical.
This means that the induced map (pull-back) $\phi^{*}$
that acts on $\Omega$ must satisfy
\begin{equation}\label{can1}
\phi^{*}\Omega=\Omega\,.
\end{equation}
As the closed canonical $2$-form $\Omega$ is locally exact
($\Omega=\d\Lambda$), and the pull-back commutes with the
exterior derivative, the difference $\phi^{*}\Lambda-\Lambda$ is closed and locally exact.
Global exactness requires an additional topological condition.
At fixed $s$ and in a local generating-function chart, we thus obtain
\begin{displaymath}
\phi^{*}\Lambda-\Lambda+\d F_{1}=0\,.
\end{displaymath}
The condition that the integrand of the extended variational
principle~\eref{principle} must remain form-invariant
is then expressed in canonical coordinates for a function
$F_{1}:\QB_{1}\times\QB_{1}\times\RB\to\RB$ as
\begin{equation}\label{can2}\fl
\sum_{i=1}^{n}p_{i}\d q^{i}-\HC\d t-H_{1}\d s=\sum_{i=1}^{n}
p^{\prime}_{i}\d q^{\prime\,i}-\HC^{\prime}\d t^{\prime}-
H_{1}^{\prime}\d s+\d F_{1}(\vecq,\vecq^{\prime},t,t^{\prime},s)\,.
\end{equation}
The requirement~\eref{can2} thus automatically ensures the
form-invariance of the canonical equations switching from
the unprimed to the primed variables.
Comparing the coordinates of the $1$-form $\d F_{1}$
\begin{displaymath}
\d F_{1}=\sum_{i=1}^{n}\left(
\frac{\partial F_{1}}{\partial q^{i}}\d q^{i}+
\frac{\partial F_{1}}{\partial q^{\prime\,i}}\d q^{\prime\,i}\right)+
\frac{\partial F_{1}}{\partial t}\d t+
\frac{\partial F_{1}}{\partial t^{\prime}}\d t^{\prime}+
\frac{\partial F_{1}}{\partial s}\d s
\end{displaymath}
with~\eref{can2}, we obtain the transformation rules ($i=1,\ldots,n$)
\begin{displaymath}\fl
p_{i}=\frac{\partial F_{1}}{\partial q^{i}}\,,\quad
p_{i}^{\prime}=-\frac{\partial F_{1}}{\partial q^{\prime\,i}}\,,\quad
\HC=-\frac{\partial F_{1}}{\partial t}\,,\quad
\HC^{\prime}=\frac{\partial F_{1}}{\partial t^{\prime}}\,,\quad
H_{1}^{\prime}=H_{1}+\frac{\partial F_{1}}{\partial s}\,.
\end{displaymath}
We immediately conclude that the extended Hamiltonian
$H_{1}$ is preserved if and only if the generating function
$F_{1}$ does not explicitly depend on $s$
\begin{equation}\label{ham1-trans}
H_{1}^{\prime}(\vecq^{\prime},\vecp^{\prime},t^{\prime},\HC^{\prime})=
H_{1}(\vecq,\vecp,t,\HC)\quad\Longleftrightarrow
\quad\partial F_{1}/\partial s=0\,.
\end{equation}
In order for the description in the new set of coordinates to be
equivalent to the original set of unprimed coordinates, the
transformation must be invertible.
For a local representation by $F_{1}$, the relevant nondegeneracy condition is
\begin{equation}\label{hessian}
\det\left(\frac{\partial^{2}F_{1}}
{\partial q^{i}\partial q^{\prime\,j}}\right)\ne0
\end{equation}
for the $(n+1)\times(n+1)$ matrix of mixed second partial derivatives, including the time coordinates.
This condition provides local invertibility and does not establish a global generating function.

With the help of the Legendre transformation
\begin{equation}\label{legendre}
F_{2}(\vecq,\vecp^{\prime},t,\HC^{\prime},s)=
F_{1}(\vecq,\vecq^{\prime},t,t^{\prime},s)+\sum_{i=1}^{n}
q^{\prime\,i}p^{\prime}_{i}-t^{\prime}\HC^{\prime}\,,
\end{equation}
an equivalent generating function $F_{2}$ can be defined that depends on
the original configuration space and the new momentum coordinates.
If we compare the coefficients pertaining to the respective
$1$-forms $\d q^{i}$, $\d p_{i}^{\prime}$, $\d t$,
$\d\HC^{\prime}$, and $\d s$, we find the following coordinate
transformation rules to apply for each index $i=1,\ldots,n$,
\begin{equation}\label{rules}\fl
p_{i}=\frac{\partial F_{2}}{\partial q^{i}}\,,\quad
q^{\prime\,i}=\frac{\partial F_{2}}{\partial p_{i}^{\prime}}\,,\quad
\HC=-\frac{\partial F_{2}}{\partial t}\,,\quad
t^{\prime}=-\frac{\partial F_{2}}{\partial \HC^{\prime}}\,,\quad
H_{1}^{\prime}=H_{1}+\frac{\partial F_{2}}{\partial s}\,.
\end{equation}
Equivalent transformation rules are induced by generating
functions
\begin{displaymath}
F_{3}(\vecq^{\prime},\vecp,t^{\prime},\HC,s)=
F_{1}(\vecq,\vecq^{\prime},t,t^{\prime},s)-\sum_{i=1}^{n}
q^{i}p_{i}+t\HC
\end{displaymath}
\begin{equation}\label{rules2}\fl
p_{i}^{\prime}=-\frac{\partial F_{3}}{\partial q^{\prime\,i}}\,,\quad
q^{i}=-\frac{\partial F_{3}}{\partial p_{i}}\,,\quad
\HC^{\prime}=\frac{\partial F_{3}}{\partial t^{\prime}}\,,\quad
t=\frac{\partial F_{3}}{\partial \HC}\,,\quad
H_{1}^{\prime}=H_{1}+\frac{\partial F_{3}}{\partial s}\,,
\end{equation}
and
\begin{displaymath}
F_{4}(\vecp,\vecp^{\prime},\HC,\HC^{\prime},s)=
F_{3}(\vecq^{\prime},\vecp,t^{\prime},\HC,s)+\sum_{i=1}^{n}
q^{\prime\,i}p_{i}^{\prime}-t^{\prime}\HC^{\prime}
\end{displaymath}
\begin{displaymath}\fl
q^{i}=-\frac{\partial F_{4}}{\partial p_{i}}\,,\quad
q^{\prime\,i}=\frac{\partial F_{4}}{\partial p_{i}^{\prime}}\,,\quad
t=\frac{\partial F_{4}}{\partial \HC}\,,\quad
t^{\prime}=-\frac{\partial F_{4}}{\partial \HC^{\prime}}\,,\quad
H_{1}^{\prime}=H_{1}+\frac{\partial F_{4}}{\partial s}\,.
\end{displaymath}
The extended Hamiltonian $H_{1}$ again obeys the passive scalar transformation law if $F_{2,3,4}$ do not explicitly depend on $s$.
The canonical equations also retain their form for explicitly $s$-dependent generators, with the additional term $\partial F/\partial s$ in the transformed Hamiltonian.
In that case the transformed physical constraint is the image of the original constraint, not generally the zero level of $H_{1}^{\prime}$.
The examples below use generators without explicit $s$ dependence.
Of course, the Hessian condition~\eref{hessian} must apply
similarly for $F_{2,3,4}$ in order to assure the invertibility
of the generated symplectic map.

The transformation rule for the conventional
(non-extended) Hamiltonians $H(\vecq,\vecp,t)$ and
$H^{\prime}(\vecq^{\prime},\vecp^{\prime},t^{\prime})$ follows finally
from the common transformation rule for the extended Hamiltonians
 $H_{1}(\vecq_{1},\vecp_{1})$ and
$H_{1}^{\prime}(\vecq_{1}^{\prime},\vecp_{1}^{\prime})$,
\begin{displaymath}
H_{1}^{\prime}=H_{1}+\frac{\partial F}{\partial s},
\end{displaymath}
with $F$ standing for one of the particular generating functions
$F_{1,2,3,4}$.
If the chosen off-surface extensions are written as $H_{1}=k(H-\HC)$ and $H_{1}^{\prime}=k^{\prime}(H^{\prime}-\HC^{\prime})$, this rule becomes
\begin{equation}\label{canham}
k^{\prime}\big[H^{\prime}(\vecq^{\prime},\vecp^{\prime},t^{\prime})-\HC^{\prime}\big]
=k\big[H(\vecq,\vecp,t)-\HC\big]+\frac{\partial F}{\partial s}\,.
\end{equation}
For an $s$-independent transformation $\phi$, the transformed constraint is initially $\mathcal C\circ\phi^{-1}=0$.
It can be solved locally as $\HC^{\prime}=H^{\prime}(\vecq^{\prime},\vecp^{\prime},t^{\prime})$ if its partial derivative with respect to $\HC^{\prime}$ does not vanish on the surface.
A nonzero factor relating this normalized constraint to $\mathcal C\circ\phi^{-1}$ is absorbed into $k^{\prime}$.
This local solvability condition also requires $t^{\prime}$ to change along the physical flow.

On the physical surface, the ratio of clock rates is the total derivative
\begin{displaymath}
\frac{\d t^{\prime}}{\d t}
=\frac{\partial t^{\prime}}{\partial t}
+\sum_i\left(\frac{\partial H}{\partial p_i}\frac{\partial t^{\prime}}{\partial q^i}
-\frac{\partial H}{\partial q^i}\frac{\partial t^{\prime}}{\partial p_i}\right)
+\frac{\partial H}{\partial t}\frac{\partial t^{\prime}}{\partial\HC}\,.
\end{displaymath}
It must not in general be replaced by $\partial t^{\prime}/\partial t$.
An equality restricted to $H-\HC=H^{\prime}-\HC^{\prime}=0$ does not by itself determine the off-surface Hamiltonians.

A useful special case is a generator $F_{2}=f_{2}(\vecq,\vecp^{\prime},t)-\HC^{\prime}T(t)$ with $T^{\prime}(t)\ne0$.
Here $t^{\prime}=T(t)$, and the energy rule gives $\HC=T^{\prime}(t)\HC^{\prime}-\partial f_{2}/\partial t$.
Consequently, the normalized Hamiltonians satisfy the exact identity
\begin{equation}\label{canham1}
\big[H^{\prime}(\vecq^{\prime},\vecp^{\prime},t^{\prime})-\HC^{\prime}\big]
\frac{\partial t^{\prime}}{\partial t}=H(\vecq,\vecp,t)-\HC\,,
\qquad t^{\prime}=T(t)\,.
\end{equation}
Equivalently, $H^{\prime}=(H+\partial f_{2}/\partial t)/T^{\prime}(t)$ after expressing the old spatial variables in the new ones.
Equation~\eref{canham1} will be used only for this time-only class of transformations.

With the set of extended canonical transformations providing
a superset of the con\-ven\-tion\-al ones, it is not astonishing
that a conventional generating function
$f_{2}(\vecq,\vecp^{\prime},t)$ can always be
reformulated as a particular extended generating function
$F_{2}(\vecq,\vecp^{\prime},t,\HC^{\prime})$ by means of
\begin{equation}\label{extension}
F_{2}(\vecq,\vecp^{\prime},t,\HC^{\prime})=
f_{2}(\vecq,\vecp^{\prime},t)-t\HC^{\prime}\,.
\end{equation}
The related transformation rules follow from Equations~\eref{rules} as
\begin{displaymath}
p_{i}=\frac{\partial f_{2}}{\partial q^{i}}\,,\quad
q^{\prime\,i}=\frac{\partial f_{2}}{\partial p_{i}^{\prime}}\,,
\quad\HC=\HC^{\prime}-\frac{\partial f_{2}}{\partial t}\,,\quad
t^{\prime}=t\,,\quad H_{1}^{\prime}=H_{1}\,.
\end{displaymath}
As the generating function from Equation~\eref{extension} does
not explicitly depend on $s$, the non-extended Hamiltonians
transform according to Eq.~(\ref{canham1}), which yields with
the above transformation rules for $\HC$ and $\HC^{\prime}$
\begin{displaymath}
H^{\prime}=H+\frac{\partial f_{2}}{\partial t}\,.
\end{displaymath}
The conventional transformations --- generated by $f_{2}$ ---
thus coincide with the particular {\em subset\/} of general
transformations generated by $F_{2}$ from Equation~\eref{extension}.
In other words, the conventional canonical transformations
distinguish themselves by the fact that the system evolution
parameter $s$ can be replaced by the time $t$ as the common
independent variable of both the original and the destination
systems.
In this respect, the transformations generated by $f_{2}$
are the time-dependent canonical transformations on the
presymplectic evolution space $T^{*}\QB\times\RB$
of definition~5.2.6 of Abraham and Marsden 1978.

According to the fourth rule of Equations~\eref{rules},
the ``extended'' generating functions $F_{2}$ in general define
non-trivial time transformations $t^{\prime}\ne t$.
As will become clear in the following example section, it is
this {\em freedom\/} to relate a given system to a destination
system at {\em different\/} instants of their respective time
scales that enables us to formulate the Lorentz transformation as
a particular canonical transformation in the extended phase space.
The generalized approach also provides useful direct transformations of suitable explicitly time-dependent systems into autonomous ones.
This property is not exclusive to transformations that change time.
For example, $H=\onehalf a(t)p^2$ is transformed into $H^{\prime}=\onehalf p^{\prime\,2}$ by the ordinary generator
\begin{displaymath}
f_{2}(q,p^{\prime},t)=qp^{\prime}
+\onehalf\left[t-\int^{t}a(\tau)\d\tau\right]p^{\prime\,2}\,.
\end{displaymath}

The extended $2$-form $\Omega$
has the highest non-vanishing power
\begin{displaymath}\fl
\Omega^{n+1}=(n+1)!\cdot (-1)^{n(n+1)/2}\cdot
\d p_{1}\wedge\cdots\wedge\d p_{n+1}\wedge\d
q^{1}\wedge\cdots\wedge\d q^{n+1}\,.
\end{displaymath}
The invariance of $\Omega$ with respect to canonical
transformations thus induces the invariance of the
extended volume form $V_{1}$
\begin{displaymath}\fl
V_{1}=\d p_{1}\wedge\cdots\wedge\d p_{n+1}\wedge\d
q^{1}\wedge\cdots\wedge\d q^{n+1}=
\d p_{1}^{\prime}\wedge\cdots\wedge\d p_{n+1}^{\prime}\wedge\d
q^{\prime\,1}\wedge\cdots\wedge\d q^{\prime\,n+1}\,.
\end{displaymath}
This is, in canonical coordinates, the extended phase-space
formulation of Liouville's theorem.

A sufficient condition for the transformed clock to be independent of the instantaneous spatial state is
\begin{equation}\label{restrict1}
\frac{\partial t^{\prime}}{\partial q^{i}}=
\frac{\partial t^{\prime}}{\partial p_{i}}=0\,.
\end{equation}
This is a useful restriction for time-only transformations of multi-particle systems, but it is not a necessary condition for every possible common evolution parameter.
A collective state-dependent clock can also parameterize all coordinates when its rate is nonzero.
Lorentz transformation of individual simultaneous particle events is a different issue, since those events need not remain simultaneous in the transformed frame.

If in addition
\begin{equation}\label{restrict2}
\frac{\partial t^{\prime}}{\partial\HC}=
\frac{\partial q^{\prime\,i}}{\partial\HC}=
\frac{\partial p^{\prime}_{i}}{\partial\HC}=0\,,
\end{equation}
then $t^{\prime}=T(t)$ and the spatial transformation is independent of $\HC$.
Locally, such a canonical transformation can be composed from a conventional time-dependent canonical transformation and a pure time scaling transformation of Section~\ref{sec:tscal}.
The Lorentz transformation does not belong to this subclass.
For a time-dependent spatial transformation, this factorization into maps should not be confused with separate preservation of both symplectic $2$-forms on a fixed product decomposition.
Mixed terms containing $\d t$ can occur.

Canonicity and the conditions above imply
\begin{equation}\label{restrict3}
\frac{\partial t^{\prime}}{\partial t}
\frac{\partial\HC^{\prime}}{\partial\HC}=1\,.
\end{equation}
Restricting the symplectic identity to fixed $t$ and $\HC$ then shows that the instantaneous spatial map preserves the conventional Liouville volume on $T^{*}\QB$.
Without these restrictions, preservation of the full extended volume does not imply preservation of every projected or fixed-coordinate subspace volume.
\section{Examples of canonical transformations in the extended phase space}
\subsection{\label{sec:lorentz}Lorentz transformation}
Here we consider two Cartesian frames of reference $(x,y,z)$ and
$(x^{\prime},y^{\prime},z^{\prime})$ that
move with respect to each other at a constant velocity $v$.
For simplicity, we first set up the coordinate system to be
aligned so that the relative motion occurs along the $x$-axis.
Under these circumstances, the $y$- and $z$-coordinates are
not affected by the Lorentz transformation ($y^{\prime}=y$,
$z^{\prime}=z$).
As this transformation necessarily involves a non-trivial
mapping of the respective time scales $t\mapsto t^{\prime}$,
it cannot be described in terms of a canonical transformation
in the conventional phase space.
Nevertheless, in the extended phase space, a generating function
$F_{2}$ exists that exactly yields the Lorentz transformation rules,
\begin{equation}\label{lorentz1}
F_{2}(x,p_{x}^{\prime},t,\HC^{\prime})=\gamma\left[
p_{x}^{\prime}(x-\beta ct)-\frac{\HC^{\prime}}{c}(ct-\beta x)\right]\,.
\end{equation}
The spectator terms $yp_y^{\prime}+zp_z^{\prime}$ are understood when the transverse canonical pairs are included.
With $c$ denoting the speed of light, the common notation is
used to express the scaled relative velocity in the
abbreviated form $\beta=v/c$.
As usual, $\gamma$ stands for the relativistic factor, defined by
$\gamma^{-2}=1-\beta^{2}$.

For the particular generating function~\eref{lorentz1}, the
general rules for a canonical trans\-for\-mation in the extended
phase space from Equations~\eref{rules} specialize to
\begin{eqnarray*}
p_{x}=\frac{\partial F_{2}}{\partial x}&=&
\gamma\left(p_{x}^{\prime}+\beta\frac{\HC^{\prime}}{c}\right)\,,
\quad\;\,\,\HC=-\frac{\partial F_{2}}{\partial t}=\gamma\left(
\HC^{\prime}+\beta c\,p_{x}^{\prime}\right),\\
x^{\prime}=\frac{\partial F_{2}}{\partial p_{x}^{\prime}}
&=& \gamma\left(x-\beta c\,t\right)\,,\qquad\quad\!
t^{\prime}=-\frac{\partial F_{2}}{\partial\HC^{\prime}}=
\gamma\left(t-\frac{\beta}{c}x\right),\\
H_{1}^{\prime}(x^{\prime},p_{x}^{\prime},t^{\prime},\HC^{\prime})
&=& H_{1}(x,p_{x},t,\HC)\,,\quad\quad
H=\gamma H^{\prime}+\beta\gamma c\,p_{x}^{\prime}.
\end{eqnarray*}
The displayed relation between $H$ and $H^{\prime}$ holds on the corresponding physical energy surfaces when the transformed constraint can be solved for $\HC^{\prime}$.
It follows from the energy transformation rule, not from~\eref{canham1}, whose time-only hypothesis is not satisfied here.
In particular, along a physical trajectory,
\begin{displaymath}
\frac{\d t^{\prime}}{\d t}=\gamma\left(1-\frac{v}{c^{2}}\frac{\d x}{\d t}\right)\,,
\qquad\frac{\partial t^{\prime}}{\partial t}=\gamma\,.
\end{displaymath}
In complex notation, the coordinate transformation rules take
on the familiar form of an orthogonal linear mapping
\begin{equation}\label{l1-rules}
\left(\begin{array}{c}x^{\prime}\\ict^{\prime}\\p_{x}^{\prime}\\
i\HC^{\prime}/c\end{array}\right)=\left(\begin{array}{cccc}
\gamma & i\beta\gamma & 0 & 0\\
-i\beta\gamma & \gamma & 0 & 0\\
0 & 0 & \gamma & i\beta\gamma\\
0 & 0 & -i\beta\gamma & \gamma\\
\end{array}\right)
\left(\begin{array}{c}x\\ict\\p_{x}\\ i\HC/c\end{array}\right)\,.
\end{equation}
We observe that the generating function \eref{lorentz1}
provides both the transformation rules for the $(x,ct)$
coordinates --- which commonly refer to the
Lorentz transformation --- and the related rules
for the conjugate coordinates momentum and energy $(p_{x},\HC/c)$.
This is not astonishing, as a canonical transformation always
maintains the symplectic structure of the Hamiltonian in question
--- which requires the transformation rules for all canonical
variables to be uniquely defined.

For a general boost direction without an additional spatial rotation,
their relative scaled velocity is expressed by the $3$-component vector
$\boldsymbol{\beta}=(\beta^{i})$.
With $\vecq=(x,y,z)=(q^{i})$ and $\vecp^{\prime}=(p_{x}^{\prime},
p_{y}^{\prime},p_{z}^{\prime})=(p_{i}^{\prime})$,
the general form of the generating function $F_{2}$ for the
Lorentz transformation is then given by
\begin{eqnarray}\label{lorentz2}\fl
F_{2}(\vecq,\vecp^{\prime},t,\HC^{\prime})=\gamma
\frac{\HC^{\prime}}{c}\left[\sum_{i=1}^{3}\beta^{i}q^{i}-ct\right]+
\sum_{i=1}^{3}p_{i}^{\prime}\left[\sum_{k=1}^{3}\left(\delta^{ik}+
(\gamma-1)\frac{\beta^{i}\beta^{k}}{{|\boldsymbol{\beta}|}^{2}}
\right)q^{k}-\gamma c\,t\beta^{i}\right]\nonumber\\
\end{eqnarray}
which simplifies to the generating function \eref{lorentz1}
for the aligned case $\beta^{1}=\beta_{x}=\beta$,
$\beta^{2}=\beta_{y}=0$, and $\beta^{3}=\beta_{z}=0$.

The generating function~\eref{lorentz2} of the Lorentz
transformation has the particular property
\begin{equation}\label{lorentzcond}
-\frac{\partial^{2}F_{2}}{\partial q^{i}\partial\HC^{\prime}}=
\frac{\partial t^{\prime}}{\partial q^{i}}=-\frac{\gamma}{c}\,\beta^{i}\,,
\end{equation}
which does not vanish for $\beta^{i}\ne0$.
For individual events that are simultaneous in the original frame, the transformed times generally differ when their positions differ.
Thus, a Lorentz boost of all such events cannot in general be interpreted as a map between two equal-time multi-particle states without additional synchronization conditions.
This relativity of simultaneity is distinct from choosing a common abstract evolution parameter.
The boost also lies outside the factorized time-only subclass of~\eref{restrict1} and~\eref{restrict2}.
The full extended Liouville volume is preserved.
Preservation of every spatial subspace volume does not follow, although the transverse canonical pairs of an $x$-directed boost are unchanged and their area forms are preserved separately.

In order to relativistically describe multi-particle systems,
Sorge \etal 1989 introduced an $8N$-dimensional phase space,
in which the positions and momenta of $N$ particles are described
as two $4$-vectors that depend on a common evolution parameter.
A more general, field-theoretical approach has been developed
by Gotay (Gotay \etal 1997) on a ``multiphase space'' that is
endowed with a ``multisymplectic'' $(n+2)$-form as the covariant
generalization of the symplectic $2$-form of Hamiltonian mechanics.

We furthermore observe that the generating function~\eref{lorentz2} does not explicitly depend on $s$.
According to~\eref{ham1-trans}, this gives $H_{1}^{\prime}(z^{\prime})=H_{1}(z)$ as a passive transformation law.
It does not establish Lorentz invariance of arbitrary dynamics.
Physical Lorentz covariance requires the specified constraint, including any external fields, to transform covariantly.

As an example, consider the non-relativistic Hamiltonian of a particle with mass $m>0$ and charge $\zeta$ within an electromagnetic field defined by the potentials $(\vecA,\phi)$,
\begin{equation}\label{fp-nl}
H_{\mathrm{NL}}(\vecq,\vecP,t)=
\frac{{\left[\vecP-\zeta\vecA(\vecq,t)\right]}^{2}}{2m}+\zeta\phi(\vecq,t)\,.
\end{equation}
This Hamiltonian is not a Lorentz scalar.
Writing $\boldsymbol\pi=\vecP-\zeta\vecA$ and $\varepsilon=\HC-\zeta\phi$, a relativistic mass-shell constraint is
\begin{displaymath}
\mathcal K=\frac{1}{2m}\left[\boldsymbol\pi^{2}
-\frac{\varepsilon^{2}}{c^{2}}+m^{2}c^{2}\right]=0\,.
\end{displaymath}
Here $\mathcal K$ is a Lorentz scalar when the electromagnetic four-potential and the canonical four-momentum are transformed consistently.
This is additional physical input, not a consequence of canonicity alone.
The expression
\begin{equation}\label{fp-l}\fl
H_{\mathrm{L}}(\vecq,\vecP,t,\HC)=\frac{1}{2m}\left[{(\vecP-\zeta\vecA)}^{2}-
\frac{{(\HC-\zeta\phi-mc^{2})}^{2}}{c^{2}}\right]+\zeta\phi+mc^{2}
\end{equation}
obeys the exact identity $H_{\mathrm L}=\HC+\mathcal K$.
Thus $H_{\mathrm L}-\HC$, rather than $H_{\mathrm L}$ itself, is the Lorentz-scalar constraint.
The notation $H_{\mathrm L}$ is retained for comparison with the non-relativistic expression.
It is not being used as a conventional coordinate-time Hamiltonian while $\HC$ remains an independent argument.

On the physical surface, imposing $H_{\mathrm L}=\HC$ gives the implicit coordinate-time Hamiltonian equation
\begin{equation}\label{fp-l2}\fl
H_{+}(\vecq,\vecP,t)=\frac{1}{2m}\left[{(\vecP-\zeta\vecA)}^{2}-
\frac{{(H_{+}-\zeta\phi-mc^{2})}^{2}}{c^{2}}\right]+\zeta\phi+mc^{2}\,.
\end{equation}
Selecting the future-directed branch, $\varepsilon>0$, yields
\begin{displaymath}
H_{+}(\vecq,\vecP,t)=\sqrt{c^{2}{\left[
\vecP-\zeta\vecA(\vecq,t)\right]}^{2}+m^{2}c^{4}}+\zeta\phi(\vecq,t)\,.
\end{displaymath}
The other algebraic branch has negative kinetic energy coordinate $\varepsilon$.
The positive branch is the familiar coordinate-time Hamiltonian obtained by solving the invariant mass-shell constraint.
It is not itself a Lorentz scalar.
For vanishing kinetic momentum and zero field, it gives the rest energy $mc^{2}$.

The distinction between a constraint and its extension away from the physical surface is visible in the exact factorization
\begin{displaymath}
\mathcal K=(H_{+}-\HC)\frac{H_{+}+\HC-2\zeta\phi}{2mc^{2}}\,.
\end{displaymath}
The Hamiltonian $\mathcal K$ generates the clock equation
\begin{displaymath}
\frac{\d t}{\d s}=-\frac{\partial\mathcal K}{\partial\HC}
=\frac{\HC-\zeta\phi}{mc^{2}}\,.
\end{displaymath}
Consequently,
\begin{displaymath}
\mathcal K=(H_{+}-\HC)\frac{\d t}{\d s}
+\frac{(H_{+}-\HC)^{2}}{2mc^{2}}\,.
\end{displaymath}
The coefficient in the exact factorization and the actual clock rate agree on the mass shell but need not agree away from it.
This is precisely the distinction described in Section~\ref{sec:principle}.
\subsection{\label{sec:tscal}Euler's time scaling transformation}
In the context of the regularizing transformation of the
three-body problem (Siegel and Moser 1971), we encounter
from heuristic reasoning a replacement of the time $t$
by a new independent variable, $t^{\prime}$, defined by
\begin{displaymath}
t^{\prime}(t)=\int_{t_{0}}^{t}\frac{\d\tau}{\xi(\tau)}\,.
\end{displaymath}
We shall see in the following that this particular
transformation constitutes a simple canonical transformation
in the extended phase space.
Namely, a canonical transformation that defines an identical mapping
in $\vecq$ and $\vecp$ but merely ``scales'' the extended variables
$t(s)$ and $\HC(s)$ is induced by the extended generating function
\begin{equation}\label{gen-tscal}
F_{2}\big(\vecq,\vecp^{\prime},t,\HC^{\prime}\big)=
\vecq\,\vecp^{\prime}-\HC^{\prime}
\int_{t_{0}}^{t}\frac{\d\tau}{\xi(\tau)}\,,
\end{equation}
with $\xi(t)$ denoting a prescribed differentiable function of time only that is nonzero on the interval considered.
The particular transformation rules emerge from the general
rules~\eref{rules} as
\begin{equation}\label{tscalrules}
q^{\prime\,i}=q^{i}\,,\quad p_{i}^{\prime}=p_{i}\,,\quad
t^{\prime}=\int_{t_{0}}^{t}\frac{\d\tau}{\xi(\tau)}\,,\quad
\HC^{\prime}=\xi(t)\,\HC\,,\quad H_{1}^{\prime}=H_{1}\,.
\end{equation}
From $H_{1}^{\prime}(\vecq_{1}^{\prime},\vecp_{1}^{\prime})=%
H_{1}(\vecq_{1},\vecp_{1})$, hence from
$H-\HC=(H^{\prime}-\HC^{\prime})\,\partial t^{\prime}/\partial t$,
we directly conclude
\begin{displaymath}
H^{\prime}\big(\vecq(t^{\prime}),\vecp(t^{\prime}),t^{\prime}\big)=
\xi(t)\,H\big(\vecq(t),\vecp(t),t\big)\,.
\end{displaymath}
Not surprisingly, the Hamiltonians $H$ and $H^{\prime}$
follow again the same transformation rule as their respective
values, $\HC$ and $\HC^{\prime}$.
Because of
\begin{displaymath}
\frac{\partial^{2}F_{2}}{\partial q^{i}\partial\HC^{\prime}}=
\frac{\partial^{2}F_{2}}{\partial p_{i}^{\prime}\partial\HC^{\prime}}=0\,,
\end{displaymath}
the transformed time $t^{\prime}$ retains the character of a
global parameter that is common to all coordinates $q^{\prime\,i}$
and $p_{i}^{\prime}$ in the transformed system.
A factorization of the extended phase space
$T^{*}\QB_{1}=T^{*}\QB\times T^{*}\RB$ is preserved, since additionally
\begin{displaymath}
\frac{\partial t^{\prime}}{\partial\HC}=
\frac{\partial q^{\prime\,i}}{\partial\HC}=
\frac{\partial p^{\prime}_{i}}{\partial\HC}=0\,.
\end{displaymath}
Because furthermore
\begin{displaymath}
\frac{\partial t^{\prime}}{\partial t}
\frac{\partial\HC^{\prime}}{\partial\HC}=1\,,
\end{displaymath}
the volume forms on the subspaces $T^{*}\QB$ and $T^{*}\RB$ are
separately conserved by means of the canonical transformation
generated by~\eref{gen-tscal}, hence
\begin{displaymath}
V=\d p_{1}\wedge\cdots\wedge\d p_{n}\wedge\d
q^{1}\wedge\cdots\wedge\d q^{n}=V^{\prime}\,,\quad
\d t\wedge\d\HC=\d t^{\prime}\wedge\d\HC^{\prime}\,.
\end{displaymath}
For a prescribed time-only function $\xi$, the formulas above define one canonical transformation on an open phase-space domain.
One may choose that function to coincide with a coordinate-dependent quantity along one previously selected trajectory.
Such a choice is then trajectory-dependent and does not by itself define a single time-only transformation for all trajectories.
A phase-space-dependent clock rate can instead be implemented directly by the Hamiltonian $H_{1}=k(H-\HC)$, with all derivatives of $k$ retained away from the constraint surface.

A simple example of how to formulate a time-scaling transformation
as an extended canonical transformation will be presented in the
following for Euler's regularization of the Kepler equation of
motion in one dimension.
In normalized variables with unit kinetic mass, the Hamiltonian of this problem is (Stiefel and Scheifele 1971)
\begin{displaymath}
H(x,p)=\onehalf p^{2}-\frac{K^{2}}{x}\,,\qquad K^{2}>0\,,\qquad x>0\,.
\end{displaymath}
Here $K^{2}$ denotes the positive gravitational coupling in the chosen normalization.
As $H$ does not explicitly depend on time, we have
$\d\HC/\d t=\partial H/\partial t=0$, hence
\begin{equation}\label{kepenergy}
\HC=\onehalf p^{2}-\frac{K^{2}}{x}={\rm const}\,.
\end{equation}
The resulting equation of motion for $x$ is obviously
singular at the point of collision at $x=0$,
\begin{displaymath}
\frac{\d^{2}x}{\d t^{2}}+\frac{K^{2}}{x^{2}}=0\,.
\end{displaymath}
It was L.~Euler who first worked out a transformation of the
time scales that regularizes this equation of motion.
First consider a prescribed clock transformation generated by $F_{2}$ from Equation~\eref{gen-tscal}, with the transformation rules~\eref{tscalrules}.
The transformed Hamiltonian $H^{\prime}$ is then
\begin{equation}\label{hamkep1}
H^{\prime}(x,p,t^{\prime})=\xi(t^{\prime})\left(\onehalf
p^{2}-\frac{K^{2}}{x}\right)\,,
\end{equation}
Here $\xi(t^{\prime})$ abbreviates the prescribed function $\xi(t(t^{\prime}))$, and the canonical trajectories are parameterized by $t^{\prime}$.
As a result of the fact that the transformation is canonical, the
form of the canonical equations is preserved in the transformed system,
\begin{displaymath}
\frac{\d x}{\d t^{\prime}}=\frac{\partial H^{\prime}}{\partial p}=
\xi(t^{\prime})\,p\,,\qquad
\frac{\d p}{\d t^{\prime}}=-\frac{\partial H^{\prime}}{\partial x}=
-\xi(t^{\prime})\,\frac{K^{2}}{x^{2}}\,.
\end{displaymath}
The equation of motion in terms of $t^{\prime}$ and the
energy conservation relation from Equation~\eref{kepenergy} follow as
\begin{equation}\label{kepeqmo}
\frac{\d^{2} x}{\d t^{\prime\,2}}-\frac{1}{\xi}
\frac{\d\xi}{\d t^{\prime}}\frac{\d x}{\d t^{\prime}}+
\frac{K^{2}\xi^{2}}{x^{2}}=0\,,\quad
{\left(\frac{\d x}{\d t^{\prime}}\right)}^{2}=
2\xi^{2}\left(\HC+\frac{K^{2}}{x}\right)\,.
\end{equation}
Along a selected trajectory, we can now choose the clock so that the prescribed rate coincides with $x(t^{\prime})$.
This determines a trajectory-dependent relation between the fictitious time $t^{\prime}$ and the physical time $t$.
In the present case we define
\begin{displaymath}
\xi(t^{\prime})\equiv x(t^{\prime})\quad\Longrightarrow\quad
t(t^{\prime})=\int_{0}^{t^{\prime}}x(\tau)\,\d\tau\,.
\end{displaymath}
The equation of motion and the energy conservation relation
from Equation~\eref{kepeqmo} now reduce to
\begin{displaymath}
\frac{\d^{2} x}{\d t^{\prime\,2}}-\frac{1}{x}
{\left(\frac{\d x}{\d t^{\prime}}\right)}^{2}+K^{2}=0\,,\quad
{\left(\frac{\d x}{\d t^{\prime}}\right)}^{2}=
2\HC\,x^{2}+2K^{2}\,x\,.
\end{displaymath}
By finally inserting the energy conservation relation
into the equation of motion, we obtain Euler's regularized
equation of motion
\begin{displaymath}
\frac{\d^{2} x}{\d t^{\prime\,2}}-2\HC\,x=K^{2}\,.
\end{displaymath}
The same regularization can be formulated for all trajectories at once by using the phase-space-dependent scaling $k=x$ in the constrained Hamiltonian,
\begin{displaymath}
\mathcal K_{\mathrm{Kep}}=x(H-\HC)
=\onehalf xp^{2}-K^{2}-x\HC\,.
\end{displaymath}
With $s$ the independent parameter, its unrestricted equations include
\begin{displaymath}
\frac{\d t}{\d s}=x\,,\qquad
\frac{\d x}{\d s}=xp\,,\qquad
\frac{\d p}{\d s}=\HC-\onehalf p^{2}\,,\qquad
\frac{\d\HC}{\d s}=0\,.
\end{displaymath}
On $\mathcal K_{\mathrm{Kep}}=0$, differentiation gives $\d^{2}x/\d s^{2}=2\HC x+K^{2}$, in agreement with Euler's equation above after identifying the evolution parameters.
Here the dependence of $k$ on $x$ is retained during differentiation.
The extra terms away from the constraint surface are required, not inconsistent with Hamilton's principle.
The equation for $x$ has a regular continuation to collision, but the physical clock rate vanishes at $x=0$ and the original momentum can diverge.
Thus the time transformation is not invertible there, and time scaling alone is not a regular canonical coordinate chart through collision.
The KS construction in Section~\ref{sec:ks} includes the additional spatial regularization.
\subsection{\label{sec:hosci}Time-dependent damped harmonic oscillator}
The time-dependent harmonic oscillator model is frequently
used as the first-order approxi\-mation for non-linear,
explicitly time-dependent Hamiltonian systems.
We shall demonstrate in the following that a system
with $n$ degrees of freedom that is confined within a time-dependent
harmonic oscillator potential and that is subject to linear
time-dependent damping forces can be mapped into a conventional
undamped time-independent harmonic oscillator system  by means
of a single canonical transformation in the extended phase space.
The Hamiltonian of the original system is given by
\begin{equation}\label{ham-tdosci}
H(\vecq,\vecp,t)=\onehalf e^{-F(t)}\vecp^{2}+
\onehalf e^{F(t)}\omega^{2}(t)\,\vecq^{2}\,.
\end{equation}
which yields the equations of motion
\begin{equation}\label{hoeqm}
\ddot{q}^{\,i}+f(t)\,\dot{q}^{i}+\omega^{2}(t)\,q^{i}=0\,,\quad
i=1,\ldots,n\,,\quad f(t)=\dot{F}(t)\,.
\end{equation}
The destination Hamiltonian $H^{\prime}$ --- with $t^{\prime}$
its independent variable --- shall be the autonomous system
\begin{equation}\label{ham-tiosci}
H^{\prime}(\vecq^{\prime},\vecp^{\prime})=
\onehalf\vecp^{\prime\,2}+
\onehalf\omega^{2}_{0}\,\vecq^{\prime\,2}\,.
\end{equation}
A family of functions
$F_{2}(\vecq,\vecp^{\prime},t,\HC^{\prime})$
that generates the mapping of the Hamiltonian~\eref{ham-tdosci}
into the Hamiltonian~\eref{ham-tiosci} has been found to be
\begin{equation}\label{gen-tdosci}
F_{2}\big(\vecq,\vecp^{\prime},t,\HC^{\prime}\big)=
\sqrt{\frac{e^{F(t)}}{\xi(t)}}\vecq\vecp^{\prime}+
\quarter e^{F(t)}\!\left[\frac{\dot{\xi}(t)}{\xi(t)}-f(t)\right]
\vecq^{2}-\HC^{\prime}\!\int_{0}^{t}\frac{\d\tau}{\xi(\tau)},
\end{equation}
with $\xi(t)>0$, for the moment, an undetermined sufficiently differentiable function of time.
Because of the quadratic dependence on the canonical coordinates,
the transformation rules~\eref{rules} yield the linear mapping
\begin{equation}\label{oscirules}
\left(\begin{array}{c}q^{\prime\,i}\\[\medskipamount] p^{\prime}_{i}\end{array}
\right)=\left(\begin{array}{cc}\sqrt{e^{F(t)}/\xi(t)} & 0\\[\medskipamount]
-\onehalf(\dot{\xi}-\xi f)\sqrt{e^{F(t)}/\xi(t)} &
\sqrt{\xi(t)/e^{F(t)}}\end{array}\right)
\left(\begin{array}{c}q^{i}\,\\[\medskipamount] p_{i}\end{array}\right).
\end{equation}
The transformations of time $t$, energy $\HC$, and extended
Hamiltonian $H_{1}$ between both systems emerge from the
generating function~\eref{gen-tdosci} as
\begin{eqnarray}\label{oscirules2}
t^{\prime}&=&\int_{0}^{t}\frac{\d\tau}{\xi(\tau)}\,,\nonumber\\
\HC^{\prime}&=&\xi\HC-\onehalf\left(\dot{\xi}-\xi f\right)\,\vecq\vecp+
\quarter e^{F(t)}\left(\ddot{\xi}-\dot{\xi}f-\xi\dot{f}\right)\,\vecq^{2}
\,,\\ H_{1}^{\prime}&=&H_{1}\nonumber.
\end{eqnarray}
We observe that the time-shift transformation between both
systems is determined by the as yet unknown function $\xi(t)$.
In any case, the transformed time $t^{\prime}$ does not depend
on individual particles coordinates, hence retains the character
of $t$ as a global parameter for all particles.
The correlation of the Hamiltonians $H$ and $H^{\prime}$ follows
from Eq.~\eref{canham1}.
With $\partial t^{\prime}/\partial t=\xi^{-1}(t)$ and inserting the
Hamiltonian~\eref{ham-tdosci}, we obtain $H^{\prime}$ as
\begin{displaymath}
H^{\prime}(\vecq^{\prime},\vecp^{\prime},t^{\prime})=
\onehalf\vecp^{\prime\,2}+\onehalf\vecq^{\prime\,2}
\left[\onehalf\xi\ddot{\xi}-\quarter{\dot{\xi}}^{2}+
\xi^{2}\left(\omega^{2}-\onehalf\dot{f}-\quarter f^{2}\right)\right]\,,
\end{displaymath}
having replaced all unprimed variables according
to the rules \eref{oscirules} and \eref{oscirules2}.
Hence, the destination Hamiltonian~\eref{ham-tiosci}
indeed emerges if we identify
\begin{equation}\label{auxosci0}
\onehalf\xi\ddot{\xi}-\quarter{\dot{\xi}}^{2}+\xi^{2}\left(\omega^{2}(t)-
\onehalf\dot{f}-\quarter f^{2}\right)=\omega_{0}^{2}=\mathrm{const}\,.
\end{equation}
The primed system's potential is not time-dependent
if and only if $\d\omega_{0}^{2}/\d t^{\prime}=0$.
This condition yields the linear third-order equation
\begin{equation}\label{auxosci}
\dddot{\xi}(t)+\dot{\xi}\left(4\omega^{2}(t)-2\dot{f}(t)-f^{2}(t)\right)+
\xi\left(4\omega\dot{\omega}-\ddot{f}-f\dot{f}\right)=0\,.
\end{equation}
Together with initial conditions, this requirement determines $\xi(t)$ and hence the time correlation $t^{\prime}(t)$ of both systems.
It is precisely the extended canonical-transformation approach
that enables us to properly adjust this time correlation.
With $\xi(t)$ a solution of~\eref{auxosci}, the Hamiltonian
$H^{\prime}$ does not depend on time explicitly.
Expressed in the unprimed coordinates, the value $\HC^{\prime}$
of $H^{\prime}$ then yields an invariant of the original
system~\eref{ham-tdosci}
\begin{equation}\label{invosci}\fl
\HC^{\prime}=\onehalf e^{-F}\xi\vecp^{2}-\onehalf\left(\dot{\xi}-
\xi f\right)\vecq\vecp+\quarter e^{F(t)}\left(\ddot{\xi}-
\dot{\xi}f-\xi\dot{f}+2\xi\omega^{2}(t)\right)\vecq^{2}=\mathrm{const}\,.
\end{equation}
In terms of $\rho(t)=\sqrt{\xi(t)}$, the invariant~\eref{invosci}
agrees with the invariant found by Leach (Leach 1978) for the
case $n=1$.
A trajectory-dependent choice of $\xi$ also recovers familiar angular-momentum-type invariants of this isotropic linear system.
We easily verify that a particular solution $\xi(t)$ of
Equation~\eref{auxosci} is given by
\begin{equation}\label{hoxi}
\xi(t)=e^{F(t)}\vecq^{2}(t)\,,
\end{equation}
of course provided that $\vecq(t)$ is a solution vector of the
equations of motion~\eref{hoeqm}.
Inserting~\eref{hoxi} and its first and second
time derivatives into~\eref{invosci}, the invariant
$\HC^{\prime}$ takes on the simple form
\begin{displaymath}
\HC^{\prime}=\vecq^{2}\vecp^{2}-{\left(\vecq\,\vecp\right)}^{2}=
\onehalf\sum_{i,j=1}^{n}{\left(p_{i}\,q^{j}-p_{j}\,q^{i}\right)}^{2}\,.
\end{displaymath}
The constancy of this sum of squares does not by itself imply constancy of its individual terms.
Their conservation follows directly from the equations of motion.
For
\begin{displaymath}
I_{i}^{j}=p_{i}q^{j}-p_{j}q^{i}\,,\qquad i,j=1,\ldots,n\,,
\end{displaymath}
substitution of $\dot q^{i}=e^{-F}p_i$ and $\dot p_i=-e^{F}\omega^{2}q^{i}$ gives $\d I_i^j/\d t=0$.
These are canonical angular-momentum-type invariants (Leach 1977).
When the coordinates describe a physical central-force problem, the corresponding mechanical angular momentum is multiplied by $e^{-F(t)}$ and need not be conserved in the presence of damping.
The choice~\eref{hoxi} depends on the selected trajectory and can vanish.
For $n=1$ the associated invariant vanishes identically.
Thus this choice need not define a nondegenerate real canonical transformation on the whole interval.

The canonical equivalence of~\eref{ham-tdosci} and \eref{ham-tiosci}
is {\em physical\/} as long as the transformation rules~\eref{oscirules}
describe the correlation of {\em real\/} particles coordinates,
hence as long as $\xi(t)>0$.
Positivity of $\xi(0)$ alone is not sufficient.
For example, $f=0$, $\omega=1$, and $\xi(t)=\cos^{2}t$ solve~\eref{auxosci}, but $\xi$ reaches zero and the constant in~\eref{auxosci0} is $\omega_{0}^{2}=0$.
Choose instead $\omega_{0}^{2}>0$.
If a smooth solution with $\xi(0)>0$ had a first zero at $t_{*}$, Equation~\eref{auxosci0} would imply
\begin{displaymath}
\omega_{0}^{2}=-\quarter\dot\xi(t_{*})^{2}\le0\,,
\end{displaymath}
which is a contradiction.
A corresponding existence construction follows by setting $\xi=\rho^{2}$ in~\eref{auxosci0},
\begin{displaymath}
\ddot\rho+\left(\omega^{2}-\onehalf\dot f-\quarter f^{2}\right)\rho
=\frac{\omega_{0}^{2}}{\rho^{3}}\,.
\end{displaymath}
This is the Ermakov-Pinney equation (Pinney 1950).
Let $u$ and $v$ be independent real solutions of its associated linear equation and let $W=u\dot v-\dot u v\ne0$.
For constants $A_{0}>0$, $C_{0}>0$, and $A_{0}C_{0}-B_{0}^{2}=\omega_{0}^{2}/W^{2}>0$, the function
\begin{displaymath}
\rho(t)=\sqrt{A_{0}u(t)^{2}+2B_{0}u(t)v(t)+C_{0}v(t)^{2}}
\end{displaymath}
is positive wherever the linear solutions are regular.
The coefficient matrix of this quadratic form is positive definite, and independent linear solutions do not vanish simultaneously.
Thus a real nondegenerate transformation to an undamped autonomous oscillator exists on intervals where the original coefficients and the transformation remain regular.
The corresponding $t^{\prime}$ interval need not cover the entire real line.
\subsection{\label{sec:genosci}General time-dependent potential}
As a generalization of the previous example, we now determine which non-linear time-dependent potentials are transformed into time-independent ones by the same scaling generator.
Let the Hamiltonian of the original system be given by
\begin{equation}\label{ham-gtdosci}
H(\vecq,\vecp,t)=\onehalf\vecp^{2}+V(\vecq,t)\,.
\end{equation}
Again, we require a destination system $H^{\prime}$ of the same
{\em form}, but with a potential $V^{\prime}$ that does {\em not\/}
explicitly depend on the system's independent variable, $t^{\prime}$,
\begin{equation}\label{ham-gtiosci}
H^{\prime}(\vecq^{\prime},\vecp^{\prime})=
\onehalf\vecp^{\prime\,2}+V^{\prime}(\vecq^{\prime})\,.
\end{equation}
The generating function~\eref{gen-tdosci} retains the quadratic momentum dependence of $H^{\prime}$ and a momentum-independent transformed potential (Struckmeier and Riedel 2002a, 2002b).
We set $F(t)=f(t)=0$, since~\eref{ham-gtdosci} does not include damping forces.
This is a scaling ansatz rather than the most general generator with these properties.
For example, spatial translations provide transformations outside this ansatz.
The transformed Hamiltonian $H^{\prime}$ is then again obtained
from the particular transformation rule from Eq.~\eref{canham1}
as $H^{\prime}=\xi(H-\HC)+\HC^{\prime}$.
Inserting the Hamiltonian $H$ of~\eref{ham-gtdosci} and the
transformation rule for $\HC^{\prime}$ of~\eref{oscirules2},
and replacing the unprimed coordinates, we find
\begin{displaymath}
H^{\prime}(\vecq^{\prime},\vecp^{\prime},t^{\prime})=
\onehalf\vecp^{\prime\,2}+\quarter\vecq^{\prime\,2}
\big(\xi\ddot{\xi}-\onehalf\dot{\xi}^{2}\big)+\xi\,
V\big(\sqrt{\xi}\,\vecq^{\prime},t\big)\,.
\end{displaymath}
Thus, a Hamiltonian $H^{\prime}$ of the form of \eref{ham-gtiosci}
turns out if the transformed potential $V^{\prime}$ is identified with
\begin{equation}\label{pot-gtiosci}
V^{\prime}\big(\vecq^{\prime},t^{\prime}\big)=\quarter\vecq^{\prime\,2}
\left(\xi\ddot{\xi}-\onehalf\dot{\xi}^{2}\right)+\xi\,V\big(\sqrt{\xi}\,
\vecq^{\prime},t\big)\,.
\end{equation}
We can now make use of the freedom to appropriately adjust
the time correlation $t^{\prime}(t)$ between the original
system~\eref{ham-gtdosci} and the destination
system~\eref{ham-gtiosci} by requiring the
new potential $V^{\prime}$ to be independent of
its time $t^{\prime}$ explicitly
\begin{equation}\label{potcond}
\frac{\partial V^{\prime}}{\partial t^{\prime}}\stackrel{!}{=}0\,.
\end{equation}
This requirement restricts both the potential and the possible functions $\xi(t)$.
It must hold for every spatial argument with one and the same $\xi$, not merely along a selected trajectory.
For the potential \eref{pot-gtiosci},
the condition~\eref{potcond} evaluates to
\begin{equation}\label{auxeq}
\dddot{\xi}\vecq^{2}+4\dot{\xi}\left[V(\vecq,t)+\onehalf
\vecq\,\frac{\partial V}{\partial\vecq}\right]+
4\xi\,\frac{\partial V}{\partial t}=0\,.
\end{equation}
Equation~\eref{auxeq} is a functional identity in $\vecq$ and $t$ when autonomy is required throughout a configuration-space domain.
Writing $\rho=\sqrt\xi>0$, Equation~\eref{pot-gtiosci} becomes
\begin{displaymath}
V^{\prime}(\vecq^{\prime},t^{\prime})
=\onehalf\rho^{3}\ddot\rho\,\vecq^{\prime\,2}
+\rho^{2}V(\rho\vecq^{\prime},t)\,.
\end{displaymath}
Consequently, the exact admissible class for the displayed generator is
\begin{equation}\label{admissible-potential}
V(\vecq,t)=\frac{1}{\rho(t)^{2}}
U\left(\frac{\vecq}{\rho(t)}\right)
-\frac{\ddot\rho(t)}{2\rho(t)}\vecq^{2}\,,\qquad \rho(t)>0\,,
\end{equation}
where $U$ is time-independent.
Direct substitution gives $V^{\prime}=U$.
An arbitrary potential need not belong to this class.
For example, in dimensionless variables $V(q,t)=q^{4}+tq^{6}$ gives
\begin{displaymath}
\dddot\xi\,q^{2}+12\dot\xi\,q^{4}
+(16t\dot\xi+4\xi)q^{6}=0\,.
\end{displaymath}
An identity in $q$ would require $\dot\xi=0$ and then $\xi=0$, so no admissible scaling of the displayed form exists for this example.
This does not rule out other canonical transformations.

For a general potential, a different statement is obtained by imposing~\eref{auxeq} only along one physical trajectory.
On intervals where $\vecq(t)^{2}\ne0$, it becomes a linear homogeneous third-order ordinary differential equation, equivalent to
\begin{equation}\label{auxsys}
\frac{\d}{\d t}\left(\begin{array}{c}\xi\\\dot{\xi}\\\ddot{\xi}
\end{array}\right)=A(t)
\left(\begin{array}{c}\xi\\\dot{\xi}\\\ddot{\xi}\end{array}\right)\,,\quad
A(t)=\left(\begin{array}{ccc}0&1&0\\0&0&1\\ -g_{1}(t)&-g_{2}(t)&0
\end{array}\right)
\end{equation}
with the coefficients $g_{1}$ and $g_{2}$ defined by
\begin{displaymath}
g_{1}(t)=\frac{4}{\vecq^{2}}
\frac{\partial V}{\partial t}\,,\qquad
g_{2}(t)=\frac{4}{\vecq^{2}}
\left[V(\vecq,t)+\onehalf\vecq\;
\frac{\partial V}{\partial\vecq}\right]\,.
\end{displaymath}
In this auxiliary construction, $g_{1}$ and $g_{2}$ are evaluated along the chosen solution $\vecq(t)$.
The system~\eref{auxsys} may be integrated together with the canonical equations.
Its coefficients, and hence its fundamental matrix, generally depend on the original initial data.
It does not impose the autonomy condition away from that trajectory.
Points where $\vecq^{2}=0$ require a separate treatment of the undivided identity~\eref{auxeq}.

Regarding the system matrix $A(t)$, we observe that its
trace is {\em always\/} zero.
Hence, the Wronski determinant of any $3\times 3$ solution
matrix $\Xi(t)$ of \eref{auxsys} is always constant, regardless
of the particular form of the system's potential $V(\vecq,t)$.
With the $3\times 3$ unit matrix as a particular initial
condition ($\Xi(0)=\Eins$), we thus obtain
\begin{equation}\label{wronski}
\Xi(t)=\left(\begin{array}{ccc}
\xi_{1}&\xi_{2}&\xi_{3}\\\dot{\xi}_{1}&\dot{\xi}_{2}&\dot{\xi}_{3}\\
\ddot{\xi}_{1}&\ddot{\xi}_{2}&\ddot{\xi}_{3}\end{array}\right)\,,
\qquad\Xi(0)=\Eins\,,\qquad\det\Xi(t)\equiv1\,.
\end{equation}
For any solution $\xi$ of this trajectory-dependent auxiliary equation, the expression suggested by~\eref{oscirules2} is constant along the selected trajectory,
\begin{equation}\label{invgosci}
I=\xi(t)\,\HC-\onehalf\dot{\xi}(t)\,
\vecq\,\vecp+\quarter\ddot{\xi}(t)\,\vecq^{2}=\mathrm{const}\,.
\end{equation}
For an admissible positive $\xi$ satisfying the identity throughout the domain, $I$ is the transformed energy $\HC^{\prime}$ of an autonomous system.
For the trajectory-dependent construction, it is instead an auxiliary conserved pairing.
Indeed, define
\begin{displaymath}
\boldsymbol y(t)=\left(\begin{array}{c}\HC\\-\onehalf\vecq\vecp\\
\quarter\vecq^{2}\end{array}\right)\,.
\end{displaymath}
The original canonical equations imply $\dot{\boldsymbol y}=-A^{T}\boldsymbol y$.
Together with $\dot\Xi=A\Xi$, this gives $\d(\Xi^{T}\boldsymbol y)/\d t=0$.
The normalization $\Xi(0)=\Eins$ therefore yields
\begin{equation}\label{invgsys}
\left(\begin{array}{c}\HC_{0}\\-\onehalf\vecq_{0}\,\vecp_{0}\\
\quarter\vecq_{0}^{2}\end{array}\right)=
\left(\begin{array}{ccc}
\xi_{1}&\dot{\xi}_{1}&\ddot{\xi}_{1}\\
\xi_{2}&\dot{\xi}_{2}&\ddot{\xi}_{2}\\
\xi_{3}&\dot{\xi}_{3}&\ddot{\xi}_{3}\end{array}\right)
{\left(\begin{array}{c}\HC\\-\onehalf\vecq\,\vecp\\
\quarter\vecq^{2}\end{array}\right)}_{t}\,.
\end{equation}
The three constant pairings in~\eref{invgsys} recover the initial values of the energy and the two scalar products.
In general the matrix must be written as $\Xi(t;z_{0})$, where $z_{0}$ denotes the initial state of the original system.
Thus
\begin{displaymath}
\boldsymbol y(t)=\Xi(t;z_{0})^{-T}\boldsymbol y(0)
\end{displaymath}
is a trajectory-dependent reconstruction identity, not a universal linear evolution law for $\boldsymbol y$.
The dependence of $\Xi$ on the initial state must be included when differentiating the actual map between different initial states.
Accordingly, $\det\Xi=1$ does not establish that this nonlinear collective map has unit Jacobian.
The fundamental solutions with $\xi_{2}(0)=\xi_{3}(0)=0$ are useful for the pairing identity but do not define admissible positive clock scalings at the initial time.

For an autonomous potential, $g_{1}=0$ and $\xi_{1}=1$ is a solution with the stated normalization.
The first component of~\eref{invgsys} then expresses the usual conservation law $\HC(t)=\HC_{0}$.
The other two components remain auxiliary reconstruction identities, not a general proof of additional independent phase-space first integrals available without solving the original dynamics.
The time dependence of the auxiliary transfer matrix may still provide information about a selected trajectory, as investigated by Struckmeier 2006.
Its relation to regularity or instability requires that trajectory dependence to be taken into account.
\subsection{\label{sec:ks}Kustaanheimo-Stiefel (KS) transformation}
A transformation to ``Kustaanheimo-Stiefel'' variables constitutes a generalization of a transformation to Levi-Civita variables (Stiefel and Scheifele 1971).
For the Kepler problem, it regularizes the equations of motion, permits a uniform treatment of elliptic, parabolic, and hyperbolic motion, and gives linear equations in the regularized spatial variables.
The spatial mapping and the replacement of physical time by a fictitious evolution parameter are separate parts of the construction.
Since four KS coordinates describe three physical spatial coordinates, the spatial mapping is not an invertible point transformation between unrestricted phase spaces of equal dimension.
Its canonical interpretation requires a bilinear momentum constraint and reduction by a one-parameter gauge symmetry (Zhao 2015).

To retain the generating-function notation, first prescribe a smooth nonvanishing function $\xi(t^{\prime})$ of time only.
Consider the extended generating expression
\begin{eqnarray}\label{genks-trans}
F_{3}\big(\vecq^{\prime},\vecp,t^{\prime},\HC\big)&=&
\left(-q_{1}^{\prime\,2}+q_{2}^{\prime\,2}+
q_{3}^{\prime\,2}-q_{4}^{\prime\,2}\right)p_{1}-
2\left(q_{1}^{\prime}q_{2}^{\prime}-
q_{3}^{\prime}q_{4}^{\prime}\right)p_{2}\nonumber\\
&&\mbox{}-2\left(q_{1}^{\prime}q_{3}^{\prime}+
q_{2}^{\prime}q_{4}^{\prime}\right)p_{3}+
\HC\int_{0}^{t^{\prime}}\xi(\tau)\d\tau\,.
\end{eqnarray}
The derivatives appearing in~\eref{rules2} define a canonical correspondence rather than an unrestricted invertible $F_{3}$ transformation.
The mapping of the four new spatial coordinates $\vecq^{\prime}=(q_{1}^{\prime},q_{2}^{\prime},q_{3}^{\prime},q_{4}^{\prime})$ into the three physical coordinates $\vecq=(q_{1},q_{2},q_{3})$ is
\begin{eqnarray*}
q_{1}&=&q_{1}^{\prime\,2}-q_{2}^{\prime\,2}-q_{3}^{\prime\,2}+q_{4}^{\prime\,2}\,,\\
q_{2}&=&2q_{1}^{\prime}q_{2}^{\prime}-2q_{3}^{\prime}q_{4}^{\prime}\,,\\
q_{3}&=&2q_{1}^{\prime}q_{3}^{\prime}+2q_{2}^{\prime}q_{4}^{\prime}\,.
\end{eqnarray*}
Appending $q_{4}=0$ would not supply an additional independent physical coordinate or make the mixed Hessian invertible.
The momentum rules obtained from~\eref{genks-trans} are
\begin{eqnarray*}
p_{1}^{\prime}&=&\,\,\hphantom{-}2q_{1}^{\prime}p_{1}+2q_{2}^{\prime}p_{2}+
2q_{3}^{\prime}p_{3}\,,\\
p_{2}^{\prime}&=&-2q_{2}^{\prime}p_{1}+2q_{1}^{\prime}p_{2}+2q_{4}^{\prime}p_{3}\,,\\
p_{3}^{\prime}&=&-2q_{3}^{\prime}p_{1}-2q_{4}^{\prime}p_{2}+2q_{1}^{\prime}p_{3}\,,\\
p_{4}^{\prime}&=&\,\,\hphantom{-}2q_{4}^{\prime}p_{1}-2q_{3}^{\prime}p_{2}+
2q_{2}^{\prime}p_{3}\,.
\end{eqnarray*}
These momenta are not independent.
They satisfy the bilinear constraint
\begin{equation}\label{ks-bilinear}
\mathcal B=-q_{4}^{\prime}p_{1}^{\prime}
+q_{3}^{\prime}p_{2}^{\prime}
-q_{2}^{\prime}p_{3}^{\prime}
+q_{1}^{\prime}p_{4}^{\prime}=0\,.
\end{equation}
To verify this and the kinetic-energy transformation, define
\begin{displaymath}
r=|\vecq|=\sum_{a=1}^{4}q_{a}^{\prime\,2}\,,\qquad
J_{ia}=\frac{\partial q_i}{\partial q_a^{\prime}}\,,\qquad
\boldsymbol v=(-q_{4}^{\prime},q_{3}^{\prime},-q_{2}^{\prime},q_{1}^{\prime})^{T}\,.
\end{displaymath}
Direct differentiation gives
\begin{displaymath}
\vecp^{\prime}=J^{T}\vecp\,,\qquad
J\boldsymbol v=0\,,\qquad
JJ^{T}=4r\mathbf 1_{3}\,.
\end{displaymath}
For $r>0$, the matrix $J$ has rank three.
Its transpose maps the physical momenta onto the subspace $\mathcal B=\boldsymbol v^{T}\vecp^{\prime}=0$.
On this subspace,
\begin{displaymath}
\vecp=\frac{J\vecp^{\prime}}{4r}\,,\qquad
\vecp^{\prime\,2}=4r\vecp^{2}\,.
\end{displaymath}
The Hamiltonian flow of $\mathcal B$ rotates the redundant KS coordinate around its circle fibre and leaves the physical variables unchanged.
Thus imposing $\mathcal B=0$ and identifying points on the same gauge orbit reduces the eight-dimensional spatial KS phase space to the six-dimensional physical phase space away from $r=0$.
The identity
\begin{displaymath}
\sum_{a=1}^{4}p_a^{\prime}\d q_a^{\prime}
=\sum_{i=1}^{3}p_i\d q_i
\end{displaymath}
on the correspondence shows that the induced reduced mapping preserves the canonical symplectic form.

For the prescribed time-only function in~\eref{genks-trans}, the remaining rules are
\begin{displaymath}
\HC^{\prime}=\HC\,\xi(t^{\prime})\,,\qquad
 t=\int_{0}^{t^{\prime}}\xi(\tau)\d\tau\,.
\end{displaymath}
In particular, the argument of $\xi$ in the energy rule is $t^{\prime}$.
Together with the spatial rules they give
\begin{displaymath}
\sum_{a=1}^{4}p_a^{\prime}\d q_a^{\prime}-\HC^{\prime}\d t^{\prime}
=\sum_{i=1}^{3}p_i\d q_i-\HC\d t\,.
\end{displaymath}
On the reduced space, this $s$-independent correspondence yields the passive law $H_{1}^{\prime}=H_{1}$.
For the normalized conventional Hamiltonians, $H^{\prime}=\xi(t^{\prime})H$ after the spatial substitutions.
This statement concerns a prescribed clock function and must not be confused with the state-dependent rescaling used for regularization.

For Kepler regularization, leave the canonical time-energy pair $(t,-\HC)$ unchanged and introduce a new evolution parameter $s$ through the phase-space scaling $k=r$.
For $H(\vecq,\vecp)=\onehalf\vecp^{2}-K^{2}/r$, the regularized Hamiltonian in the KS variables is
\begin{equation}\label{ks-hamiltonian}
\mathcal K_{\mathrm{KS}}=r(H-\HC)
=\frac18\sum_{a=1}^{4}p_a^{\prime\,2}
-\HC\sum_{a=1}^{4}q_a^{\prime\,2}-K^{2}\,.
\end{equation}
The equality with $r(H-\HC)$ holds on the spatial correspondence for $r>0$.
The polynomial expression on the right supplies a smooth regularized extension.
Physical initial data satisfy both $\mathcal K_{\mathrm{KS}}=0$ and $\mathcal B=0$.
Since $\{\mathcal B,\mathcal K_{\mathrm{KS}}\}=0$, the bilinear constraint is preserved.
An arbitrary multiple of $\mathcal B$ could be added to the Hamiltonian to change the gauge representative without changing the reduced physical motion.
Choosing that gauge multiplier to vanish, the canonical equations are
\begin{displaymath}
\frac{\d q_a^{\prime}}{\d s}=\frac14p_a^{\prime}\,,\qquad
\frac{\d p_a^{\prime}}{\d s}=2\HC q_a^{\prime}\,,\qquad
\frac{\d t}{\d s}=r\,,\qquad
\frac{\d\HC}{\d s}=0\,.
\end{displaymath}
Hence
\begin{equation}\label{ks-linear}
\frac{\d^{2}q_a^{\prime}}{\d s^{2}}-\frac{\HC}{2}q_a^{\prime}=0\,,
\qquad a=1,\ldots,4\,.
\end{equation}
Negative, zero, and positive $\HC$ give oscillatory, linear, and exponential solutions, respectively.
The regularized equations are nonsingular at $\vecq^{\prime}=0$, but the reconstruction of physical time and the inverse physical coordinate map are singular there.
Thus the regularization extends the dynamics through collision without making the original physical coordinates into an invertible canonical chart at collision.

\section{Conclusions}
With the present paper, we have presented a constrained formulation of classical Hamiltonian dynamics on the symplectic extended phase space.
The time $t(s)=q^{n+1}(s)$ and its conjugate momentum $-\HC(s)=p_{n+1}(s)$ are treated as independent canonical variables before the physical energy constraint is imposed.
With $\omega=\sum_i\d p_i\wedge\d q^i$ on $T^{*}\QB$, the extended symplectic form on $T^{*}\QB_{1}$ is $\Omega=\omega-\d\HC\wedge\d t$.
Its nondegeneracy gives the usual correspondence between Hamiltonian functions and vector fields on the extended phase space.

Hamilton's variational principle with an independent multiplier yields the constraint $H-\HC=0$ and the clock equation $\d t/\d s=k$.
After fixing the parameterization, a smooth nonvanishing phase-space function $k$ gives the Hamiltonian $H_{1}=k(H-\HC)$.
Its unrestricted equations contain derivatives of $k$ multiplied by the constraint.
On the physical surface they reduce to the conventional canonical equations expressed in the parameter $s$.
The choice $k=1$ and nonconstant phase-space choices are both consistent.
The singular Legendre transformation of the homogeneous extended Lagrangian yields the same constraint, not a unique extension of the Hamiltonian away from it.

Extended canonical transformations preserve the symplectic form and the form of Hamilton's equations.
For a generating function without explicit $s$ dependence, the extended Hamiltonian obeys the passive scalar transformation law $H_{1}^{\prime}(z^{\prime})=H_{1}(z)$.
An explicitly $s$-dependent generator contributes $\partial F/\partial s$ to the Hamiltonian transformation law.
The transformation of a general clock involves a total derivative along the motion.
The simplified formula involving $\partial t^{\prime}/\partial t$ applies to the time-only transformations identified in Section~\ref{sec:extended}.
Conventional time-preserving canonical transformations form a particular subclass, and the extended Liouville volume is preserved by every extended canonical map.

The Lorentz transformation provides an example outside the time-only subclass.
Its canonicity does not make arbitrary dynamics Lorentz-invariant.
For a charged relativistic particle, the invariant object is the mass-shell constraint, from which the coordinate-time square-root Hamiltonian is obtained by selecting an energy branch.
The latter Hamiltonian is not itself a Lorentz scalar.

Prescribed time-only scalings can be implemented as finite canonical transformations.
The coordinate-dependent clocks used in regularization are instead represented directly by scaled constraints.
Euler's regularized Kepler equation follows from the choice $k=x$.
The KS example additionally requires the bilinear constraint and its gauge reduction.
Together with $k=r$, it yields a polynomial regularized Hamiltonian and linear equations in the four KS coordinates.
At collision, the regularized dynamics remains well defined although the physical clock and coordinate reconstruction are singular.

The displayed oscillator generator maps the time-dependent damped harmonic oscillator to an autonomous undamped oscillator when the auxiliary scaling is chosen appropriately.
The condition $\omega_{0}^{2}>0$ supplies a positive nondegenerate scaling on regular intervals through the Ermakov-Pinney construction.
For non-linear potentials, autonomy under the same generator requires the specific potential class~\eref{admissible-potential}.
For a general potential, the auxiliary third-order equation remains useful along a selected trajectory, but its transfer matrix depends on the initial state.
Its unit determinant describes that auxiliary linear system, not a universal unit-Jacobian map of the original collective variables.

The symplectic extended phase space and its Poisson bracket therefore give a useful framework for time transformations and constrained dynamics while retaining the distinction between canonical coordinate changes, clock choices, and physical invariance.
\appendix
\section{\label{sec:appa}Extended Lagrangian description}
\subsection{Extended Euler-Lagrange equations}
A time-independent Lagrangian $L$ is defined as the mapping
of the tangent bundle $T\QB$ into $\RB$.
If the Lagrangian $L$ is explicitly time-dependent, then
its domain is $T\QB\times\RB$, with $\RB$ denoting the time axis.
In local coordinates $(q^{i},\dot{q}^{i})$, the actual system
path $(q^{i}(t),\dot{q}^{i}(t))\subset T\QB$ is given as the
solution of the variational problem
\begin{displaymath}
\delta\int_{t_{1}}^{t_{2}}L\left(q^{1},\ldots,q^{n},
\dot{q}^{1},\ldots,\dot{q}^{n},t\right)\d t\stackrel{!}{=}0\,.
\end{displaymath}
The variational integral can be expressed equivalently in
parametric form if one replaces the time $t$ as the independent
variable with a new system evolution parameter, $s$.
With
\begin{displaymath}
q^{n+1}=t\,,\qquad\dot{q}^{i}=\frac{\d q^{i}/\d s}{\d q^{n+1}/\d s}\,,
\end{displaymath}
we obtain (Lanczos 1949, Arnold 1989)
\begin{equation}\label{extendedprinciple}\fl
\delta\int_{s_{1}}^{s_{2}}L\left(q^{1},\ldots,q^{n+1},
\frac{\d q^{1}/\d s}{\d q^{n+1}/\d s},\ldots,
\frac{\d q^{n}/\d s}{\d q^{n+1}/\d s}\right)
\frac{\d q^{n+1}}{\d s}\d s\stackrel{!}{=}0\,.
\end{equation}
On the open part of $T\QB_{1}$ where $\d t/\d s\ne0$, the integrand of~\eref{extendedprinciple} defines the extended Lagrangian $L_{1}$,
\begin{equation}\label{extlagdef}\fl
L_{1}\left(\vecq_{1},\frac{\d\vecq_{1}}{\d s}\right)=
L\left(q^{1},\ldots,q^{n+1}, \frac{\d q^{1}/\d s}
{\d q^{n+1}/\d s},\ldots,\frac{\d q^{n}/\d s}
{\d q^{n+1}/\d s}\right)\frac{\d q^{n+1}}{\d s}\,,
\end{equation}
with $\vecq_{1}=(\vecq,t)\in\QB\times\RB=\QB_{1}$
the extended configuration space vector.
The local coordinate representation of the actual system path
$(q^{i}(s),\d q^{i}/\d s)\subset T\QB_{1}$ is now given as
the solution of the variational problem
\begin{equation}\label{principle2}
\delta\int_{s_{0}}^{s_{1}}L_{1}\left(\vecq_{1}(s),
\frac{\d\vecq_{1}(s)}{\d s}\right)\d s\stackrel{!}{=}0\,.
\end{equation}
As in the case of the conventional variational problem
with a Lagrangian $L(\vecq,\dot{\vecq})$, we find that
\eref{principle2} is globally fulfilled if the
extended set of Euler-Lagrange equations is satisfied,
\begin{equation}\label{ext-eleq}
\frac{\partial L_{1}}{\partial\vecq_{1}}-\frac{\d}{\d s}
\left(\frac{\partial L_{1}}{\partial(\d\vecq_{1}/\d s)}\right)=0\,.
\end{equation}
The following identities are readily derived from \eref{extlagdef}
\begin{eqnarray}
\quad\frac{\partial L_{1}}{\partial\vecq}&=&
\frac{\d t}{\d s}\frac{\partial L}{\partial\vecq}\,,\qquad\qquad
\frac{\partial L_{1}}{\partial (\d\vecq/\d s)}=
\frac{\partial L}{\partial\dot{\vecq}}\,,\\
\frac{\partial L_{1}}{\partial t}&=&
\frac{\d t}{\d s}\frac{\partial L}{\partial t}
\,,\qquad\qquad\;
\frac{\partial L_{1}}{\partial (\d t/\d s)}=L-\dot{\vecq}\,
\frac{\partial L}{\partial\dot{\vecq}}\,,\label{backtra-b}
\end{eqnarray}
which make it possible to rewrite \eref{ext-eleq}
in terms of the conventional Lagrangian $L$
\begin{equation}\label{ext-eleq1}\fl
\frac{\d t(s)}{\d s}\left[\frac{\partial L}{\partial\vecq}-\frac{\d}{\d t}
\left(\frac{\partial L}{\partial\dot{\vecq}}\right)\right]=0\,,\qquad
\frac{\d\vecq(s)}{\d s}\left[\frac{\partial L}{\partial\vecq}-
\frac{\d}{\d t}\left(\frac{\partial L}{\partial\dot{\vecq}}\right)\right]=0\,.
\end{equation}
For $\d t/\d s\ne0$, both equations~\eref{ext-eleq1} are fulfilled if and only if the equations in brackets ---
the conventional Euler-Lagrange equations --- are satisfied.
Thus, the extended set of Euler-Lagrange equations~\eref{ext-eleq}
is equivalent to the conventional set and does {\em not\/}
provide an additional equation of motion for $t=t(s)$.
This corresponds to the freedom in choosing the multiplier in the constrained Hamiltonian formulation.
A prescribed clock fixes that freedom without introducing new physical motion.
The passage from $L_{1}$ to a Hamiltonian description must take account of the singularity of its velocity Hessian.
\subsection{\label{sec:appb}Singular Legendre transformation of the extended Lagrangian $L_{1}$}
Replacing all derivatives $\d q^{i}/\d s$ with $a\,\d q^{i}/\d s$, $a>0$, we realize that $L_{1}$ from Equation~\eref{extlagdef} is a
homogeneous form of first order in the $n+1$ variables
$\d q^{1}/\d s,\ldots,\d q^{n+1}/\d s$.
Hence, Euler's theorem on homogeneous functions yields the identity
(Lanczos 1949)
\begin{equation}\label{lagid}
L_{1}\equiv\sum_{i=1}^{n+1}\frac{\partial L_{1}}
{\partial(\d q^{i}/\d s)}\frac{\d q^{i}}{\d s}\,.
\end{equation}
Differentiation of Euler's identity with respect to a velocity gives
\begin{displaymath}
\sum_{j=1}^{n+1}
\frac{\partial^{2}L_{1}}{\partial(\d q^{i}/\d s)\partial(\d q^{j}/\d s)}
\frac{\d q^{j}}{\d s}=0\,.
\end{displaymath}
Thus the extended velocity Hessian is singular.
Assuming the original spatial Legendre transformation is regular, the image of the extended Legendre map has the single energy constraint described below.
This is a local statement on a regular spatial Legendre branch.

For the indices $i=1,\ldots,n$, the partial derivatives of
$L_{1}$ along the fibres $\d q^{i}/\d s$ define the generalized
canonical momenta $p_{i}$,
\begin{equation}\label{canmom}
\frac{\partial L_{1}}{\partial(\d q^{i}/\d s)}\equiv
\frac{\partial L}{\partial\dot{q}^{i}}\equiv p_{i}\,,\quad i=1,\ldots,n\,.
\end{equation}
The partial derivative of $L_{1}$ with respect to $\d q^{n+1}/\d s$
follows from its definition in \eref{extlagdef} as
\begin{equation}\label{indnp1}
\frac{\partial L_{1}}{\partial(\d q^{n+1}/\d s)}\equiv
L-\sum_{i=1}^{n}p_{i}\dot{q}^{i}\equiv -H(\vecq,\vecp,t)\,.
\end{equation}
Inserting \eref{canmom} and \eref{indnp1} into the identity
\eref{lagid}, the extended Lagrangian $L_{1}$ takes on the form
\begin{equation}\label{lagid1}
L_{1}\equiv\sum_{i=1}^{n}p_{i}\frac{\d q^{i}}{\d s}-
H(\vecq,\vecp,t)\frac{\d q^{n+1}}{\d s}\,.
\end{equation}
The Legendre energy of $L_{1}$ is
\begin{displaymath}
E_{L_{1}}=\sum_{i=1}^{n+1}p_i\frac{\d q^i}{\d s}-L_{1}\,.
\end{displaymath}
Euler's identity~\eref{lagid} implies that this energy vanishes identically on the velocity domain.
Inserting~\eref{lagid1} expresses the same fact as
\begin{equation}\label{ham1a}
E_{L_{1}}=\big[H(\vecq,\vecp,t)+p_{n+1}\big]
\frac{\d t}{\d s}=0
\quad\hbox{on the Legendre image}\,.
\end{equation}
Equation~\eref{indnp1} therefore specifies the image of the singular Legendre map by
\begin{equation}\label{nonid}
p_{n+1}=-\HC\,,\qquad
\HC=H(\vecq,\vecp,t)\quad\hbox{on }\Sigma\,.
\end{equation}
The second equality is a constraint on the image, not an identity between independent functions throughout $T^{*}\QB_{1}$.
Since the Legendre map is singular, the identically vanishing Legendre energy cannot be interpreted as an unconstrained Hamiltonian generating the original motion on the entire extended cotangent bundle.

A total Hamiltonian with an independent multiplier is instead
\begin{equation}\label{ham1b}
H_{1}=k(s)\big[H(\vecq,\vecp,t)-\HC\big]\,,\qquad
H_{1}\big|_{\Sigma}=0\,.
\end{equation}
Its variational principle gives both the constraint and the relation $\d t/\d s=k(s)$, as in Section~\ref{sec:principle}.
After prescribing a phase-space scaling function, one obtains~\eref{ham1}, with the additional off-surface derivative terms displayed in Section~\ref{sec:caneq}.
The function $H_{1}$ can have nonzero derivatives on $\Sigma$ even though its restriction to $\Sigma$ vanishes.
There is therefore no conflict between the vanishing Legendre energy and nontrivial constrained Hamiltonian dynamics.
The constraint term is retained when the canonical variables are varied independently in the Hamiltonian action.
The singular Legendre transformation fixes the constraint but does not select a unique off-surface extension or a unique time parameterization.
\ackn
The author is indebted to C.~Riedel (GSI) for his essential
contributions to this work.
\References
\item[]Abraham R and Marsden J E 1978 {\it Foundations of Mechanics\/} 2nd edn
(Boulder, CO: Westview Press)
\item[]Arnold V I 1989 {\it Mathematical Methods of Classical Mechanics\/} 2nd edn
(New York: Springer)
\item[]Asorey M, Cari\~{n}ena J F and Ibort L A 1983
{\it J.~Math.~Phys.\/} {\bf 24} 2745
\item[]Cari\~{n}ena J F, Ibort L A, Gomis J and Rom\'an-Roy N 1987
{\it Nuovo Cimento B\/} {\bf 98} 172--196
\item[]Cari\~{n}ena J F, Ibort L A and Lacomba E A 1988
{\it Celes.~Mech.} {\bf 42} 201--13
\item[]Frankel Th 2001
{\it The Geometry of Physics\/} (Cambridge: Cambridge University Press)
\item[]Gotay M J 1982 {\it Proc.~Am.~Math.~Soc.} {\bf 84} 111
\item[]Gotay M J, Isenberg J, Marsden J E, and Montgomery R 1997
{\it Momentum Maps and Classical Relativistic Fields:
Part I. Covariant Field Theory\/} online at
http://www.math.hawaii.edu/\~{}gotay/GIMMsyI.pdf
\item[]Jos\'e J V and Saletan E J 1998 {\it Classical Dynamics\/}
(Cambridge: Cambridge University Press)
\item[]Kuwabara R 1984 {\it Rep.~Math.~Phys.}~{\bf 19} 27--38
\item[]Lanczos C 1949 {\it The Variational Principles of Mechanics\/}
(Toronto, Ontario: University of Toronto Press)
Reprint 1986 4th edn (New York: Dover Publications)
\item[]Leach P G L 1977 {\it J.~Math.~Phys.} {\bf 18} 1608
\item[]Leach P G L 1978 {\it SIAM J.~Appl.~Math.} {\bf 34} 496
\item[]Lichtenberg A J and Lieberman M A 1992
{\it Regular and Chaotic Motion\/} (New York: Springer)
\item[]Marsden J E and Ratiu T S 1999 {\it Introduction to Mechanics
and Symmetry\/} (New York: Springer)
\item[]Pinney E 1950 {\it Proc.~Am.~Math.~Soc.} {\bf 1} 681
\item[]Siegel C L and Moser J K 1971 {\it Lectures on
Celestial Mechanics\/} (Berlin: Springer)
\item[]Sorge H, St\"ocker H and Greiner W 1989 {\it Nucl.~Phys.~A}
{\bf 498} 567c
\item[]Stiefel E L and Scheifele G 1971 {\it Linear and Regular
Celestial Mechanics\/} (Berlin: Springer)
\item[]Struckmeier J and Riedel C 2001 {\it Phys.~Rev.~E\/} {\bf 64} 026503
\item[]Struckmeier J and Riedel C 2002a {\it Ann.~Phys.~Lpz.} {\bf 11} 15--38
\item[]Struckmeier J and Riedel C 2002b {\it Phys.~Rev.~E\/} {\bf 66} 066605
\item[]Struckmeier J 2006 {\it Phys.~Rev.~E\/} {\bf 74} 026209
\item[]Stump D R 1998 {\it J.~Math.~Phys.} {\bf 39} 3661
\item[]Synge J L 1960 {\it Encyclopedia of Physics\/} Vol~3/1 ed S Fl\"ugge
(Berlin: Springer)
\item[]Szebehely V 1967 {\it Theory of Orbits\/} (New York: Academic Press)
\item[]Thirring W 1977 {\it Lehrbuch der Mathematischen Physik\/} (Wien: Springer)
\item[]Tsiganov A V 2000 {\it J.~Phys.~A: Math.~Gen.} {\bf 33} 4169--82
\item[]Wodnar K 1995 Symplectic mappings and Hamiltonian systems
{\it Perturbation theory and chaos in nonlinear dynamics with emphasis
to celestial mechanics\/} ed J Hagel, M Cunha and R Dvorak R
(Portugal: Universidade da Madeira, Funchal)
\item[]Zhao L 2015 {\it Regular and Chaotic Dynamics\/} {\bf 20} 19--36
\endrefs
\end{document}